\documentclass[aps,prl,reprint,groupedaddress]{revtex4-2}
\usepackage{xcolor}
\usepackage{graphicx}
\usepackage{amssymb} %maths
\usepackage{amsmath} %maths
\usepackage[english]{babel}
\usepackage{epstopdf, epsfig}

\usepackage[version=4]{mhchem}
\usepackage{siunitx,color}

\usepackage[colorinlistoftodos]{todonotes}

\usepackage{xcolor}

\begin{document}

\title{Elastic snap-through instabilities are governed by geometric symmetries}

\author{Basile Radisson and Eva Kanso}

\affiliation{Department of Aerospace and Mechanical Engineering, University of Southern California, Los Angeles, CA 90089-1191, USA}

\date{\today}

\begin{abstract}
Many elastic structures exhibit rapid shape transitions between two possible equilibrium states: umbrellas become inverted in strong wind and hopper popper toys jump when turned inside-out. This snap-through is a general motif for the storage and rapid release of elastic energy, and it is exploited by many biological and engineered systems from the Venus flytrap to mechanical metamaterials. Shape transitions are known to be related to the type of bifurcation the system undergoes, however, to date, there is no general understanding of the mechanisms that select these bifurcations. Here we analyze numerically and analytically two systems proposed in recent literature in which an elastic strip, initially in a buckled state, is driven through shape transitions by either rotating or translating its boundaries. We show that the two systems are mathematically equivalent, and identify three cases that illustrate the entire range of transitions described by previous authors. Importantly, using reduction order methods, we establish the nature of the underlying bifurcations and explain how these bifurcations can be predicted from geometric symmetries and symmetry-breaking mechanisms, thus providing universal design rules for elastic shape transitions.
\end{abstract}


\maketitle
Bistability and snap-through transitions are key phenomena in many biological~\cite{forterre2005,smith2011} and manmade~\cite{pandey2014,gomez2017} systems. Bistability refers to a system with two stable equilibrium states. Snap-through occurs when a system is in an equilibrium state that becomes unstable or suddenly disappears, as a control parameter is varied. Familiar examples range from the Venus flytrap~\cite{forterre2005} to children's toys~\cite{pandey2014} and ancient catapults~\cite{soedel1979}. Mechanical metamaterials, whose behavior is governed by their geometric structure rather than elastic properties, can be designed to exploit these instabilities to induce shape transitions and switch between multiple modes of functionality~\cite{silverberg2014}.

\begin{figure*}[!t]
	\centering
	\includegraphics[width =\linewidth]{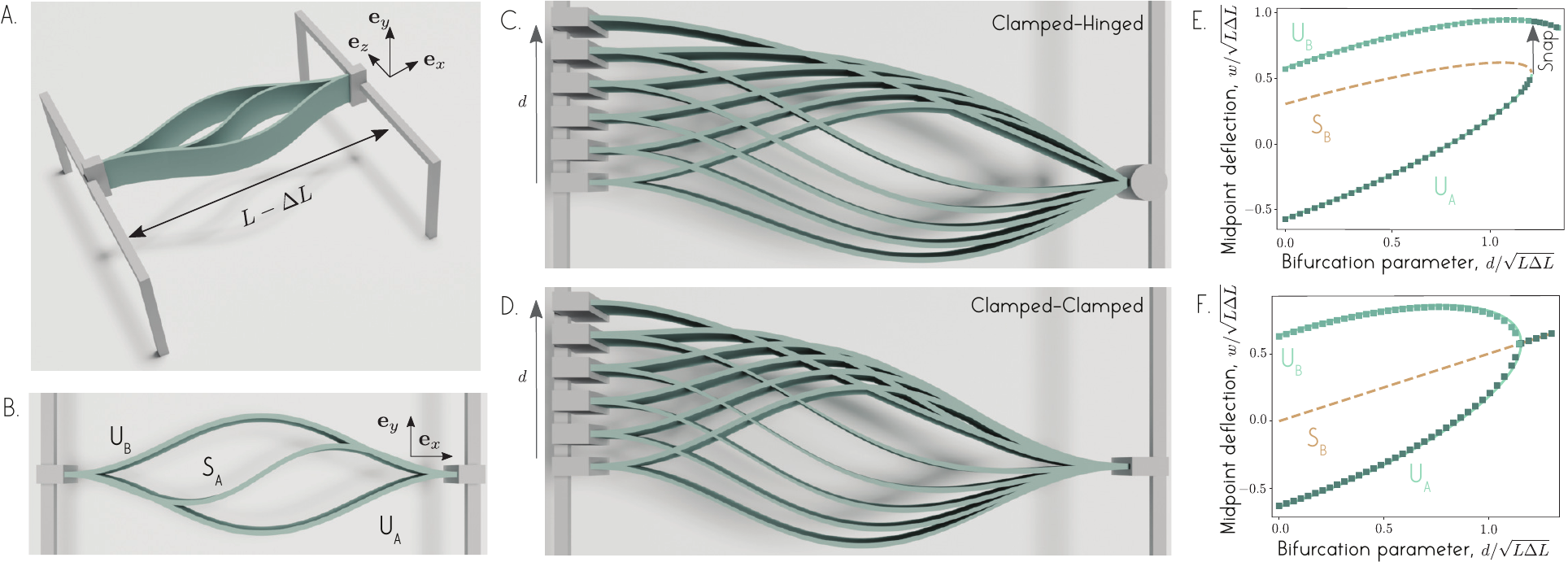}
	\caption{
	\footnotesize{(A,B) \textbf{Elastic buckled strip} with clamped-clamped boundary conditions exhibits  two symmetric stable equilibria U\textsubscript{A} and U\textsubscript{B}, and pairs of unstable equilibria of alternating symmetry at increasing energy levels, S\textsubscript{A} and S\textsubscript{B} denoting the first unstable pair  (S\textsubscript{B} not shown). (C-F) \textbf{Actuation of buckled strip} 
	by (quasi-statically) translating its left end by a distance $d$ leads to loss of bistability and a shape transition that depends on BCs: (C,E) CH strip exhibits a violent snap-through, (D,F) the transition in the CC strip  is smooth.  (A-D) 3D computer graphics rendering of the Cosserat numerical simulations. (E,F) midpoint deflection $w/\sqrt{L\Delta L}$ versus bifurcation parameter $d/\sqrt{L\Delta L}$. In all figures, (green) square markers represent data obtained based on the Cosserat rod theory. Solid (green) and dashed (brown) lines represent, respectively, stable and unstable branches obtained from the Euler beam model. 
	}
	}
	\label{fig:boundaryTranslation}
\end{figure*}

\begin{figure*}[!t]
	\centering
	\includegraphics[scale = 1]{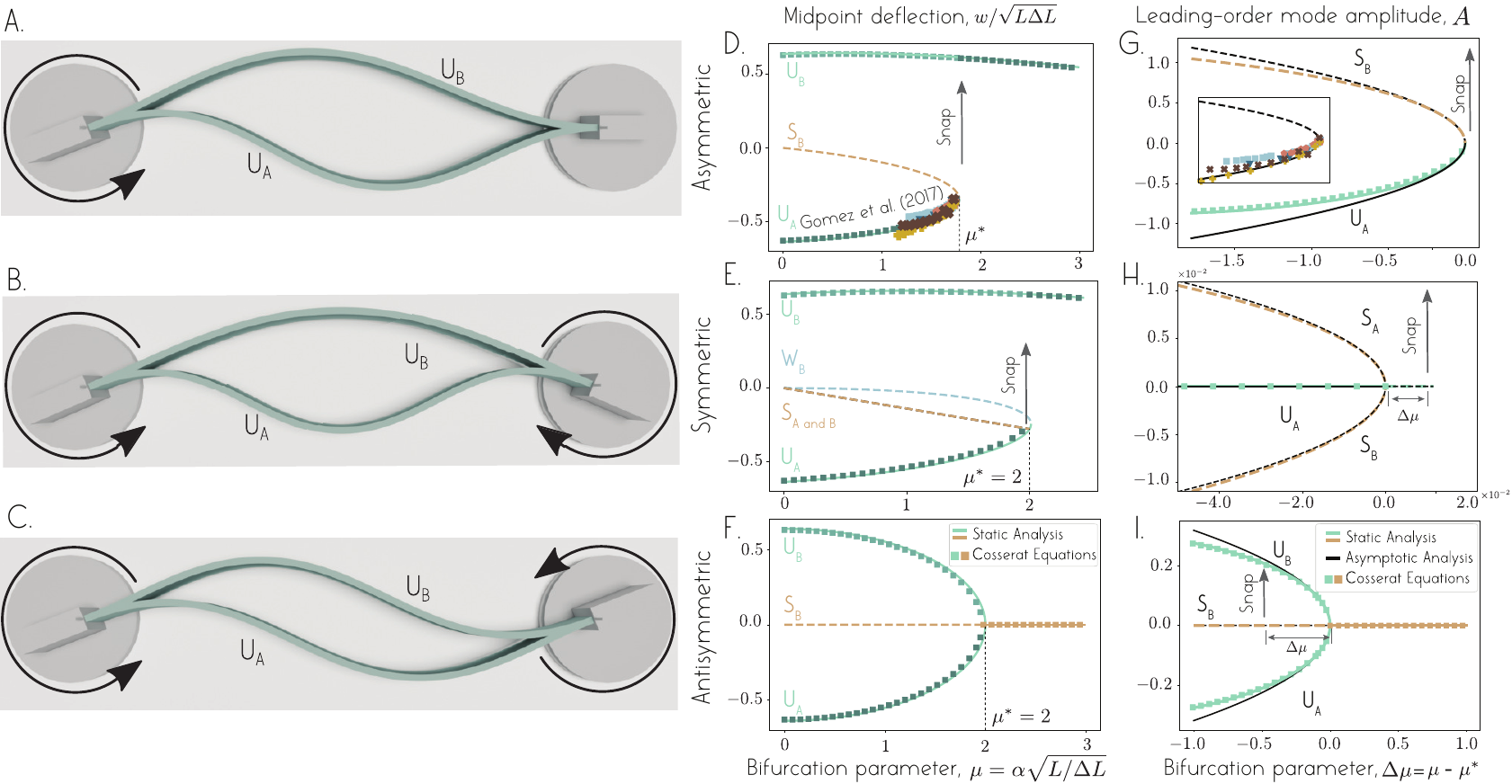}
	\caption{\footnotesize{\textbf{Boundary actuation and bifurcation diagrams} from (quasi-statically) rotating one or both clamped ends in (A)  asymmetric, (B) symmetric and (C) antisymmetric fashion. 
   (D-F) Midpoint deflection as a function of bifurcation parameter $\mu = \alpha \sqrt{L/\Delta L}$. Experimental data from \cite{gomez2017} are superimposed.
	(G-I) Asymptotic analysis near the bifurcation point $\mu^\ast$ gives access to normal forms describing the amplitude $A(t)$ of the leading order mode. Bifurcation diagrams of the normal forms (black lines) agree quantitatively with data obtained from the Euler beam equations (green and brown lines) and Cosserat simulations (green and brown square markers), and experimental data.}}
	\label{fig:boundaryRotation}
\end{figure*}

Elastic strips of length $L$, whose ends are first brought together by a distance $\Delta L$ to cause the strip to buckle into one of two stable shapes (Fig.~\ref{fig:boundaryTranslation}A, Movie S1), then driven by boundary actuation, provide an intuitive system to demonstrate shape transitions (Figs.~\ref{fig:boundaryTranslation}  and~\ref{fig:boundaryRotation}, Movies S2 and S3) \cite{gomez2017,sano2018}. 
Starting from the Euler-buckled strip with clamped-clamped (CC) BCs, when both ends are rotated symmetrically and held at a non-zero angle $\alpha$,  one equilibrium takes an ‘inverted’ shape while the other maintains its ‘natural’  shape. A larger rotation causes the inverted shape to snap to the natural shape.  Rotating only one end also creates a violent snap-through, albeit of different character~\cite{gomez2017}. 
A clamped-hinged (CH) strip with the hinged end free to rotate in place and the clamped end sheared by a distance $d$ in the direction transverse to the buckled shape exhibits snap-through~\cite{sano2018}. 
A similar set-up with CC BCs leads to graceful merging of the two equilibrium states. 

Despite the relative simplicity of realizing these transitions experimentally~\cite{gomez2017,sano2018}, 
an understanding of how shape transitions are selected remains lacking.
In a beautiful analysis, \cite{gomez2017} showed that snap-through in asymmetric BCs arises from a saddle-node bifurcation and argued that in the case of symmetric BCs, it results from a subcritical pitchfork bifurcation, without explaining what leads to this change in the character of the bifurcation as BCs change.
In~\cite{sano2018}, the authors alluded to similarities 
between their system and that of~\cite{gomez2017}.
However, to date, no general theory exists for designing systems that achieve or avoid a specific type of transition.  
Here, we combine numerical and analytical methods to reveal the mechanisms governing shape transitions in boundary-actuated elastic strips, and provide a rigorous proof that the two systems in~\cite{sano2018,gomez2017} are equivalent. 
Importantly, to predict the type of bifurcation and establish design rules for creating a desired shape transition, we show that these transitions are governed by geometric symmetries.

Symmetry is one of the most fundamental concepts in physics. 
Symmetries shape the energy landscape and govern the equilibrium configurations the system can adopt. 
Broken symmetries are often invoked to explain transitions in a range of physical systems from condensed matter physics \cite{chaikin1995} to quantum field theory \cite{peskin2018}, turbulence theory \cite{frisch1996}, fluid dynamics \cite{crawford1991}, biological locomotion \cite{michelin2010, tjhung2012}, and combustion phenomena \cite{joulin1998}. 
Simple one-dimensional (1D) examples from bifurcation theory show that a broken symmetry can turn a graceful pitchfork bifurcation into a violent saddle-node bifurcation (SI, \S S1), \cite{strogatz1994}.
Extending this understanding to infinite-dimensional systems is challenging to researchers and educators alike. The understanding we develop for elastic strips could thus serve as an educational tool to illustrate the role of symmetry-breaking in the bifurcation of continuum systems.

%-------------------
%-------------------

We investigate the bifurcation behavior of the elastic strips introduced in~\cite{sano2018,gomez2017} numerically (Figs~\ref{fig:boundaryTranslation}--\ref{fig:boundaryRotation}), by leveraging the three-dimensional (3D) Cosserat theory~\cite{cosserat1909}, and its discrete counterpart, the Discrete Elastic Rod~\cite{gazzola2018} (SI, \S S2). To establish bifurcation diagrams and carry out asymptotic analysis, we also analyze the strip's behavior in the limit of small deflection $w(x,t)$, with $-L/2<x<L/2$, based on the Euler-Bernoulli Beam theory (\cite{pandey2014,gomez2017}, and SI, \S S2),
%------
\begin{equation}
\rho b h \dfrac{\partial^2 w}{\partial t^2} + B \dfrac{\partial^4 w}{\partial x^4}  + F\dfrac{\partial^2 w}{\partial x^2}  = 0.
\label{eq:Euler}
\end{equation}
%------
The material properties of the strip are denoted by $\rho$ (density), $b$ (width), $h$ (thickness), and $B = E bh^3/12$ (bending stiffness, with $E$ the Young’s modulus). The applied compressive load is denoted by $F$. In this limit, the inextensibility condition gives rise to the nonlinear constraint equation
%----
\begin{equation}
\int_{-L/2}^{L/2} \left(\dfrac{\partial w}{\partial x}\right )^2  dx = 2 \Delta L .
\label{eq:Constraint}
\end{equation}
%-----

The Euler-buckled strip (Fig.~\ref{fig:boundaryTranslation}A,B) admits an infinite family of static equilibria that come in pairs, ordered by increasing value of elastic bending energy  $\mathcal{E}_b$ (SI, \S S4). We refer to members of the same pair as \textit{twin solutions}. 
The fundamental buckling mode, i.e., lowest energy level,
corresponds to two stable U-shape equilibria (U\textsubscript{A} and U\textsubscript{B}). Higher modes are unstable and alternate between odd and even harmonics. 
The first unstable mode gives rise to a twin of S-shape equilibria labeled S\textsubscript{A} and S\textsubscript{B}.

Through systematic numerical experiments, we investigate how boundary actuation modifies the U\textsubscript{A} and U\textsubscript{B} equilibria.
In Fig.~\ref{fig:boundaryTranslation}, we control the transverse distance $d$ at the clamped end of the CH and CC strip (SI, \S S5). 
In Fig.~\ref{fig:boundaryRotation},  we control the rotation at one or both ends of the CC strip by specifying the tangent direction (angle $\alpha$) at the boundaries (SI, \S S6).  The control parameters are varied incrementally starting from the twin solutions U\textsubscript{A,B}, allowing the elastic strip to reach mechanical equilibrium at each increment. In Fig.~\ref{fig:boundaryTranslation}E,F and Fig.~\ref{fig:boundaryRotation}D-F,  we plot the strip's midpoint deflection $w$, normalized by the length scale $\sqrt{L\Delta L}$, as a function of the non-dimensional control parameters $d/ \sqrt{L\Delta L}$ and $\alpha \sqrt{L/\Delta L}$, respectively. 
Bistability is lost beyond a certain threshold in all cases, but the character of this transition depends on boundary actuation.
Asymmetric and symmetric rotations cause snap-through from the inverted (U\textsubscript{A}) to the natural (U\textsubscript{B}) shape, as does transverse shearing of the CH strip. The dynamic evolution of the strip differs during snapping:  the displacement of the midpoint grows quadratically in time in the asymmetric case, while  it grows exponentially in time in the symmetric case~\cite{radisson2022PRE}.
Antisymmetric rotations and transverse shearing  of the CC strip induce graceful merging of the equilibrium shapes U\textsubscript{A,B}. These findings are consistent with experimental observations~\cite{gomez2017, sano2018}, and agree quantitatively with~\cite{gomez2017}. 

%--------
%---------

To understand the mechanisms leading to the similarities and differences in these shape transitions,
we solved Eqs.~(1-2) to arrive at analytic expressions for the infinite set of twin equilibria for each type of boundary actuation (SI, \S S5 and \S S6), and we assessed their linear stability subject to small perturbations (SI, \S S3). This analysis matches quantitatively the numerical solutions in Figs.~\ref{fig:boundaryTranslation}E,F and~\ref{fig:boundaryRotation}D-F for small $\Delta L$, and shows that, depending on the type of boundary actuation, the stable equilibrium U\textsubscript{A} that is energetically unfavored by the boundary actuation must collide with one or both unstable S\textsubscript{A,B} equilibria at the shape transition.

Importantly, the similarity of the bifurcation diagrams in Figs.~\ref{fig:boundaryTranslation} and~\ref{fig:boundaryRotation} is not a coincidence. 
We proved, by introducing a frame of reference attached to the line connecting the strip's endpoints (SI, \S S8), that transverse shearing of the strip is equivalent to rotation of its boundaries.
Hereafter, we focus on the strip actuated by rotating its endpoints with $\mu = \alpha \sqrt{L/\Delta L}$ as the bifurcation parameter in discussing geometric symmetries and the role they play in selecting the type of bifurcation underlying a shape transition.

%----------
\begin{figure*}[!t]
	\centering
	\includegraphics[scale =1]{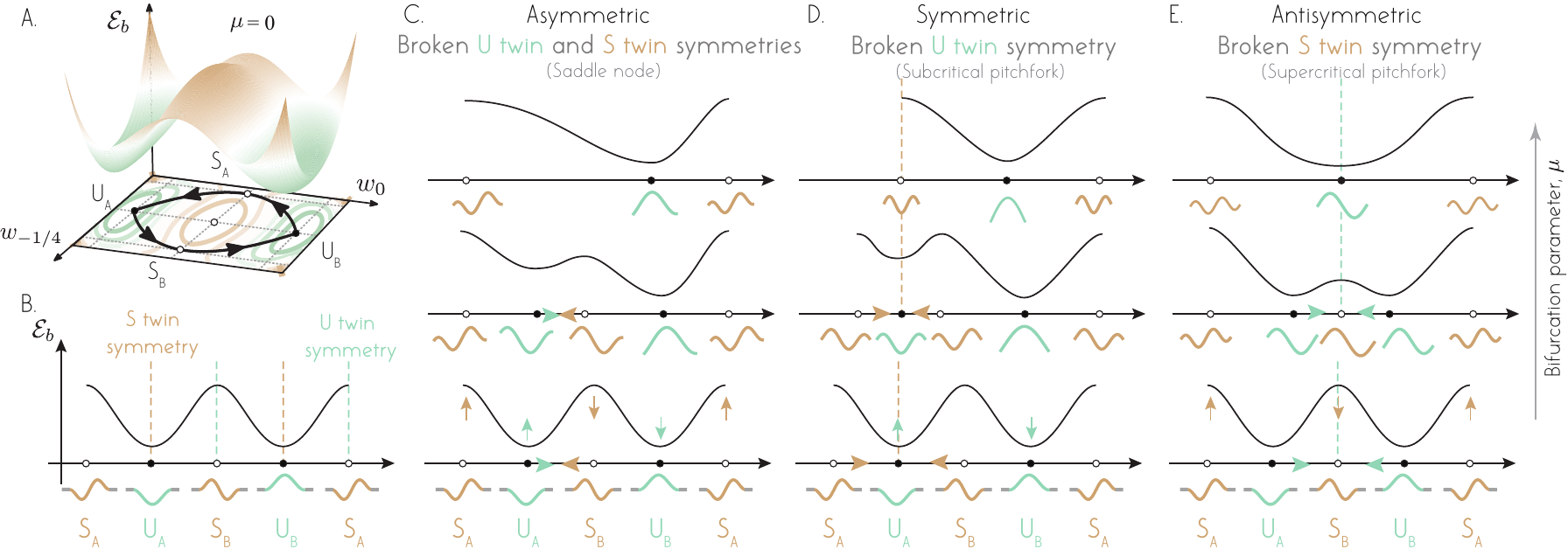}
	\caption{\footnotesize{ (A) \textbf{Energy landscape}  at $\mu=0$: two potential wells at the two stable equilibria U\textsubscript{A,B} separated by lowest energy barriers at the first pair of unstable equilibria S\textsubscript{A,B}. The two paths connecting U\textsubscript{A} to U\textsubscript{B} via either S\textsubscript{A} or S\textsubscript{B} constitutes the energetically cheapest routes to pass from  U\textsubscript{A} to U\textsubscript{B}. (B)  1D periodic representation of energy landscape. (C-E) Rotating one or both of the boundaries reshapes the energy landscape: breaking both U- and S-twin symmetries leads to a saddle-node bifurcation; breaking either 
 U- or S-twin symmetry leads to a pitchfork bifurcation.}}
	\label{fig:Energy}
\end{figure*}
%----------

So which symmetries matter? Three symmetries are important 
and best introduced  in the context of the Euler-buckled strip at $\mu =0$:
top-bottom reflection ($w\to-w$), left-right reflection  ($x\to-x$), and $\pi$-rotation ($w\to-w$ and $x\to-x$). 
Eqs.~(\ref{eq:Euler}-\ref{eq:Constraint}) are invariant under all three transformations (SI, \S S3). Because the state of the system is infinite dimensional, we calculate the bending energy $\mathcal{E}_b = (EI/2)\int_{-L/2}^{L/2} (\partial^2 w/\partial x^2)^2 dx$ at U\textsubscript{A,B} and S\textsubscript{A,B} and depict the energy landscape semi-schematically on a reduced 2D space consisting of the deflection $w$ evaluated at the strip's mid- and quarter-length (Fig.~\ref{fig:Energy}A; SI, \S S10). 
In Fig.~\ref{fig:Energy}B, we unfold the energy landscape along the closed black curve connecting the U- and S-shapes.  This representation highlights two important properties at $\mu=0$: the minimum energy barrier (difference in $\mathcal{E}_b$ between S\textsubscript{A,B} and U\textsubscript{A,B}) that the strip needs to overcome in order to undergo a shape transition from U\textsubscript{A} to U\textsubscript{B}, and the geometric symmetries that map U\textsubscript{A} to U\textsubscript{B} and S\textsubscript{A} to S\textsubscript{B}, and vice-versa. 
Specifically, the left-right symmetry maps each U-solution to itself and the top-bottom and $\pi$-rotation symmetries map a U-solution to its twin, whereas the $\pi$-rotation symmetry maps each S-solution to itself and the top-bottom and left-right symmetries map an S-solution to its twin.
Hereafter, we refer to the $\pi$-rotation  that maps the U-twin shapes to one another as the \textit{U-twin symmetry} and the left-right reflection that maps the S-twin shapes to one another as the \textit{S-twin symmetry}. The type of shape transition the system undergoes for $\mu\neq 0$ is directly related to which twin symmetry gets broken by boundary actuation.

Asymmetric boundary actuation breaks both U- and S-twin symmetries. It requires U\textsubscript{A} to bend more than U\textsubscript{B} and S\textsubscript{A} to bend more than S\textsubscript{B}, thus increasing the bending energy of U\textsubscript{A} and S\textsubscript{A} and decreasing that of U\textsubscript{B} and S\textsubscript{B} (Fig. \ref{fig:Energy}C). This causes  U\textsubscript{A} and S\textsubscript{B} to monotonically approach each other until they merge and suddenly vanish. The system must jump to U\textsubscript{B}.
Symmetric actuation breaks the U-twin symmetry but conserves the S-twin symmetry. It requires U\textsubscript{A} to bend more than U\textsubscript{B} but it equally affects S\textsubscript{A} and S\textsubscript{B}. Thus,
S\textsubscript{A} and S\textsubscript{B} remain energetically equivalent while the energetic state of U\textsubscript{A} increases and approaches that of S\textsubscript{A} and S\textsubscript{B} until they all merge in a single unstable equilibrium (Fig. \ref{fig:Energy}D), leaving the system no option but to jump to  U\textsubscript{B}.
%in a subcritical pitchfork bifurcation 
 Antisymmetric actuation conserves the U-twin symmetry but not the S-twin symmetry: U\textsubscript{A} and U\textsubscript{B}  remain energetically equivalent while S\textsubscript{A}  bends more than S\textsubscript{B}; U\textsubscript{A} and U\textsubscript{B}  monotonically approach S\textsubscript{B} until they all gracefully merge in a single stable equilibrium 
 %in a supercritical pitchfork bifurcation 
 (Fig. \ref{fig:Energy}E).

 This intuitive understanding of geometric symmetries is substantiated by extending the asymptotic analysis of \cite{gomez2017} to derive normal forms near the shape transition at $\mu^\ast$.
We set $\mu = \mu^\ast + \Delta \mu$ with $\Delta \mu \ll 1$, and introduce the dimensionless variables $X = x/L$, $W = w/\sqrt{L\Delta L}$, $W_{\textrm{eq}}^\ast = w_{\textrm{eq}}^\ast/\sqrt{L\Delta L}$, $T= t \sqrt{B/\rho h L^4}$ and $\Lambda^2 = FL^2/B$. 
To analyze the dynamic of the strip near the bifurcation, we define a slow time scale $\tau=\Delta \mu^\textrm{a}T$, and expand the dynamic state of the strip in powers of $\Delta \mu$ \cite{gomez2017, radisson2022PRE},
%----
\begin{equation}
\begin{split}
W(X,\tau) & \!=\! {W}_\textrm{eq}^*(X)+\Delta\mu^{\textrm{b}} {W}_1(X, \tau)
+ \text{h.o.t.}
%+O(\Delta \mu^{3\textrm{b}})
,\\[2mm]
\Lambda(\tau)&\!=\!\Lambda_\textrm{eq}^*+\Delta\mu^\textrm{c} \Lambda_0(\tau)\!
+ \text{h.o.t}.
\end{split}
%O(\Delta \mu^{3\textrm{b}}).
\label{eq:expansion}
\end{equation}
%----
Here, the values of $\textrm{a}$,  $\textrm{b}$, and $\textrm{c}$ depend on the intrinsic properties of the system.  In~\cite{radisson2022PRE}, we
present a systematic approach to calculate them. 
We find that, 
for the asymmetric BCs, $\textrm{a}=1/4$,  $\textrm{b}=\textrm{c}=1/2$
as postulated in~\cite{gomez2017}, whereas for the symmetric and antisymmetric BCs, $\textrm{a}=\textrm{b}=1/2$, and $\textrm{c}=1$.
We substitute $\textrm{a}$,  $\textrm{b}$, and $\textrm{c}$ into~\eqref{eq:expansion} and write 
$\Delta \mu^\textrm{b}W_0 = {A}(T)\Phi_0(X)$, where $\Phi_0(X)$ is the shape of the leading order mode and ${A}(T)$ its unscaled amplitude.
We arrive at a reduced form for each boundary actuation (see \cite{radisson2022PRE}). For the asymmetric BCs, the normal form obtained in \cite{gomez2017} is representative of a saddle node bifurcation
%-----
\begin{equation}
\frac{d^2 {A}}{dT^2}= a_{1,\textrm{asym}} \Delta \mu   +a_{2, \textrm{asym}}{A}^2,
\label{eq:saddlenodeCanonical}
\end{equation}
%-----
where $a_{1,\textrm{asym}}$ and $a_{2, \textrm{asym}}$ are positive constants (explicit expressions in \cite{gomez2017}). For the  symmetric and antisymmetric BCs, we obtain a normal  form representative of a pitchfork bifurcation  (explicit expressions of  $c_{1,(\cdot)}$ and $c_{2,(\cdot)}$  in \cite{radisson2022PRE}), 
%----
\begin{equation}
	\dfrac{d^2 {A}}{dT^2}= b_{1,(\cdot)} \Delta\mu {A}+b_{2, (\cdot)} {A}^3.
	\label{eq:pitchforkCanonical}
\end{equation}
%----------
For the symmetric case,  the coefficients $b_{1,\textrm{sym}}$ and $b_{2,\textrm{sym}}$ are positive, and the cubic term is destabilizing (subcritical pitchfork), whereas for the antisymmetric case, the coefficients $b_{1,\textrm{anti}}$ and $b_{2,\textrm{anti}}$ are negative and the cubic term is stabilizing (supercritical pitchfork).

Bifurcation analysis of~\eqref{eq:saddlenodeCanonical} and~\eqref{eq:pitchforkCanonical} recapitulates the results in Fig.~\ref{fig:Energy}.
For $\Delta \mu <0$, \eqref{eq:saddlenodeCanonical} admits a stable equilibrium (representing U\textsubscript{A})  and an unstable equilibrium (representing S\textsubscript{B}) that collide and annihilate at 
$\Delta \mu =0$ (Fig.~\ref{fig:boundaryRotation}G). As U\textsubscript{A} vanishes, the strip is forced to snap to U\textsubscript{B} (not represented in the reduced form).
For $\Delta \mu <0$,~\eqref{eq:pitchforkCanonical} admits three equilibria. In the symmetric case, these equilibria represent U\textsubscript{A}, S\textsubscript{A}, and S\textsubscript{B}, that merge at $\Delta \mu = 0$ (Fig.~\ref{fig:boundaryRotation}H). U\textsubscript{A} becomes unstable and the strip is forced to snap to U\textsubscript{B}.  In the antisymmetric case, the three equilibria represent U\textsubscript{A}, U\textsubscript{B}, and S\textsubscript{B}. They merge at $\Delta \mu = 0$  (Fig.~\ref{fig:boundaryRotation}I). The simultaneous shape change from U\textsubscript{A} and U\textsubscript{B} to S\textsubscript{B} is graceful.

To quantitatively compare this asymptotic analysis to the data in Fig.~\ref{fig:boundaryRotation}D-E, we calculated the amplitude $A$ directly from data (SI, \S S9) and plotted the results in Fig.~\ref{fig:boundaryRotation}G-I as a function of the distance from the bifurcation $\Delta \mu$, measured from the respective $\mu^\ast$ value. We observe good agreement (near $\mu^\ast$) with the bifurcation diagrams of the normal forms (black lines). Notably, the reduced forms capture correctly, not only the static shape bifurcations, but also the dynamics of snapping near these bifurcations ~\cite{gomez2017, radisson2022PRE}.

\begin{figure}[!t]
	\centering
	\includegraphics[width =\linewidth]{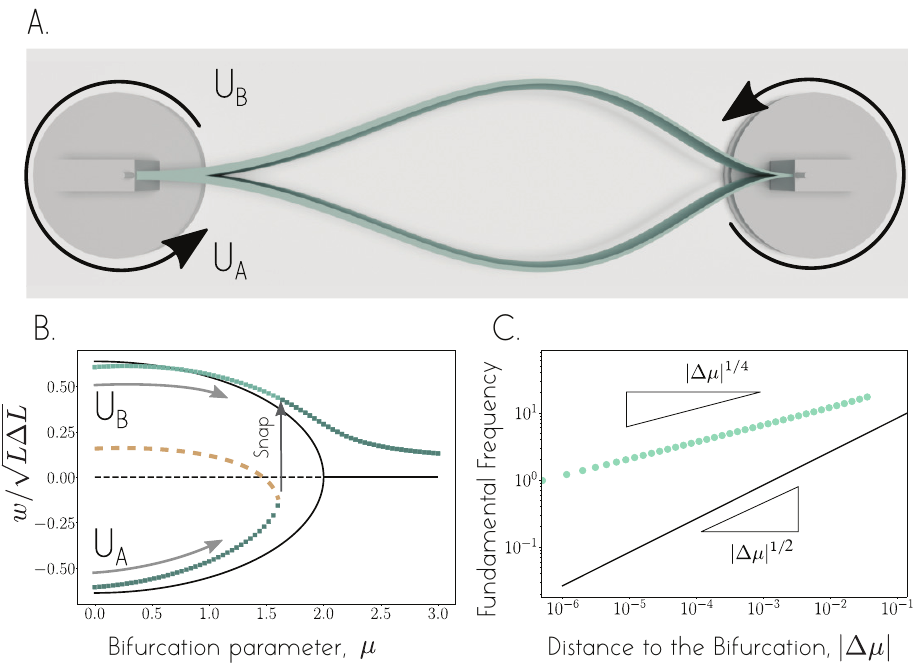}
	\caption{\footnotesize{(A) \textbf{Tapered elastic strip} under antisymmetric boundary rotation. (B)  Midpoint deflection (green symbols) versus bifurcation parameter exhibits snap-through as opposed  to the graceful merging of a homogeneous strip subject to the same actuation (black lines). (C) Critical slowing down near the bifurcation scales as $(\Delta \mu)^{1/4}$ as in the case of a saddle-node.}}
\label{fig:taperedBeam}
\end{figure}

The normal forms in~\eqref{eq:saddlenodeCanonical} and~\eqref{eq:pitchforkCanonical} provide the backbone for plotting the energy landscapes in Fig.~\ref{fig:Energy}D-F, which exhibit all the features of the rigorous bifurcation analysis (SI, \S S10).
Importantly, the well known symmetry breaking mechanism that turns a pitchfork into a saddle node bifurcation \cite{strogatz1994} (SI, \S S1), appears here, in an infinite dimensional system, governing elastic transitions. This intuitive yet universal understanding of elastic instabilities based on symmetries of the Euler-buckled strip provides powerful tools for diagnostics and design. It helps explain the force hysteresis observed in~\cite{sano2018} (SI, \S S7).
%provides a powerful design tool for elastic snap-through instabilities, and
It can also help design programmable meta-materials with tunable bistability and rapid (algebraic or exponential) actuation capabilities.
For buckled elastic strip, clamped at both ends and driven via antisymmetric rotations, to undergo a non-linear snap-through, we must break the U-twin symmetry. This can be achieved by using a strip with geometric or material heterogeneity, such as a geometrically-tapered strip instead of a homogeneous strip (Fig.~\ref{fig:taperedBeam}, SI, \S S11).  Future work will consider extensions of this analysis to elastic shells and origami-based structures~\cite{reid2017}.

\bibliography{referencesPRL}

%apsrev4-2.bst 2019-01-14 (MD) hand-edited version of apsrev4-1.bst
%Control: key (0)
%Control: author (8) initials jnrlst
%Control: editor formatted (1) identically to author
%Control: production of article title (0) allowed
%Control: page (0) single
%Control: year (1) truncated
%Control: production of eprint (0) enabled
\begin{thebibliography}{19}%
\makeatletter
\providecommand \@ifxundefined [1]{%
 \@ifx{#1\undefined}
}%
\providecommand \@ifnum [1]{%
 \ifnum #1\expandafter \@firstoftwo
 \else \expandafter \@secondoftwo
 \fi
}%
\providecommand \@ifx [1]{%
 \ifx #1\expandafter \@firstoftwo
 \else \expandafter \@secondoftwo
 \fi
}%
\providecommand \natexlab [1]{#1}%
\providecommand \enquote  [1]{``#1''}%
\providecommand \bibnamefont  [1]{#1}%
\providecommand \bibfnamefont [1]{#1}%
\providecommand \citenamefont [1]{#1}%
\providecommand \href@noop [0]{\@secondoftwo}%
\providecommand \href [0]{\begingroup \@sanitize@url \@href}%
\providecommand \@href[1]{\@@startlink{#1}\@@href}%
\providecommand \@@href[1]{\endgroup#1\@@endlink}%
\providecommand \@sanitize@url [0]{\catcode `\\12\catcode `\$12\catcode
  `\&12\catcode `\#12\catcode `\^12\catcode `\_12\catcode `\%12\relax}%
\providecommand \@@startlink[1]{}%
\providecommand \@@endlink[0]{}%
\providecommand \url  [0]{\begingroup\@sanitize@url \@url }%
\providecommand \@url [1]{\endgroup\@href {#1}{\urlprefix }}%
\providecommand \urlprefix  [0]{URL }%
\providecommand \Eprint [0]{\href }%
\providecommand \doibase [0]{https://doi.org/}%
\providecommand \selectlanguage [0]{\@gobble}%
\providecommand \bibinfo  [0]{\@secondoftwo}%
\providecommand \bibfield  [0]{\@secondoftwo}%
\providecommand \translation [1]{[#1]}%
\providecommand \BibitemOpen [0]{}%
\providecommand \bibitemStop [0]{}%
\providecommand \bibitemNoStop [0]{.\EOS\space}%
\providecommand \EOS [0]{\spacefactor3000\relax}%
\providecommand \BibitemShut  [1]{\csname bibitem#1\endcsname}%
\let\auto@bib@innerbib\@empty
%</preamble>
\bibitem [{\citenamefont {Forterre}\ \emph {et~al.}(2005)\citenamefont
  {Forterre}, \citenamefont {Skotheim}, \citenamefont {Dumais},\ and\
  \citenamefont {Mahadevan}}]{forterre2005}%
  \BibitemOpen
  \bibfield  {author} {\bibinfo {author} {\bibfnamefont {Y.}~\bibnamefont
  {Forterre}}, \bibinfo {author} {\bibfnamefont {J.~M.}\ \bibnamefont
  {Skotheim}}, \bibinfo {author} {\bibfnamefont {J.}~\bibnamefont {Dumais}},\
  and\ \bibinfo {author} {\bibfnamefont {L.}~\bibnamefont {Mahadevan}},\
  }\bibfield  {title} {{\selectlanguage {english}\bibinfo {title} {How the
  venus flytrap snaps}},\ }\href {https://doi.org/10.1038/nature03185}
  {\bibfield  {journal} {\bibinfo  {journal} {Nature}\ }\textbf {\bibinfo
  {volume} {433}},\ \bibinfo {pages} {421–425} (\bibinfo {year}
  {2005})}\BibitemShut {NoStop}%
\bibitem [{\citenamefont {Smith}\ \emph {et~al.}(2011)\citenamefont {Smith},
  \citenamefont {Yanega},\ and\ \citenamefont {Ruina}}]{smith2011}%
  \BibitemOpen
  \bibfield  {author} {\bibinfo {author} {\bibfnamefont {M.}~\bibnamefont
  {Smith}}, \bibinfo {author} {\bibfnamefont {G.}~\bibnamefont {Yanega}},\ and\
  \bibinfo {author} {\bibfnamefont {A.}~\bibnamefont {Ruina}},\ }\bibfield
  {title} {{\selectlanguage {english}\bibinfo {title} {Elastic instability
  model of rapid beak closure in hummingbirds}},\ }\href
  {https://doi.org/10.1016/j.jtbi.2011.05.007} {\bibfield  {journal} {\bibinfo
  {journal} {Journal of Theoretical Biology}\ }\textbf {\bibinfo {volume}
  {282}},\ \bibinfo {pages} {41–51} (\bibinfo {year} {2011})}\BibitemShut
  {NoStop}%
\bibitem [{\citenamefont {Pandey}\ \emph {et~al.}(2014)\citenamefont {Pandey},
  \citenamefont {Moulton}, \citenamefont {Vella},\ and\ \citenamefont
  {Holmes}}]{pandey2014}%
  \BibitemOpen
  \bibfield  {author} {\bibinfo {author} {\bibfnamefont {A.}~\bibnamefont
  {Pandey}}, \bibinfo {author} {\bibfnamefont {D.~E.}\ \bibnamefont {Moulton}},
  \bibinfo {author} {\bibfnamefont {D.}~\bibnamefont {Vella}},\ and\ \bibinfo
  {author} {\bibfnamefont {D.~P.}\ \bibnamefont {Holmes}},\ }\bibfield  {title}
  {\bibinfo {title} {Dynamics of snapping beams and jumping poppers},\ }\href
  {https://doi.org/10.1209/0295-5075/105/24001} {\bibfield  {journal} {\bibinfo
   {journal} {EPL (Europhysics Letters)}\ }\textbf {\bibinfo {volume} {105}},\
  \bibinfo {pages} {24001} (\bibinfo {year} {2014})}\BibitemShut {NoStop}%
\bibitem [{\citenamefont {Gomez}\ \emph {et~al.}(2017)\citenamefont {Gomez},
  \citenamefont {Moulton},\ and\ \citenamefont {Vella}}]{gomez2017}%
  \BibitemOpen
  \bibfield  {author} {\bibinfo {author} {\bibfnamefont {M.}~\bibnamefont
  {Gomez}}, \bibinfo {author} {\bibfnamefont {D.}~\bibnamefont {Moulton}},\
  and\ \bibinfo {author} {\bibfnamefont {D.}~\bibnamefont {Vella}},\ }\bibfield
   {title} {\bibinfo {title} {Critical slowing down in purely elastic
  ‘snap-through’ instabilities},\ }\href
  {https://doi.org/10.1038/nphys3915} {\bibfield  {journal} {\bibinfo
  {journal} {Nature Physics}\ }\textbf {\bibinfo {volume} {13}},\ \bibinfo
  {pages} {142–145} (\bibinfo {year} {2017})}\BibitemShut {NoStop}%
\bibitem [{\citenamefont {Soedel}\ and\ \citenamefont
  {Foley}(1979)}]{soedel1979}%
  \BibitemOpen
  \bibfield  {author} {\bibinfo {author} {\bibfnamefont {W.}~\bibnamefont
  {Soedel}}\ and\ \bibinfo {author} {\bibfnamefont {V.}~\bibnamefont {Foley}},\
  }\bibfield  {title} {{\selectlanguage {english}\bibinfo {title} {Ancient
  catapults}},\ }\href {https://doi.org/10.1038/scientificamerican0379-150}
  {\bibfield  {journal} {\bibinfo  {journal} {Scientific American}\ }\textbf
  {\bibinfo {volume} {240}},\ \bibinfo {pages} {150–161} (\bibinfo {year}
  {1979})}\BibitemShut {NoStop}%
\bibitem [{\citenamefont {Silverberg}\ \emph {et~al.}(2014)\citenamefont
  {Silverberg}, \citenamefont {Evans}, \citenamefont {McLeod}, \citenamefont
  {Hayward}, \citenamefont {Hull}, \citenamefont {Santangelo},\ and\
  \citenamefont {Cohen}}]{silverberg2014}%
  \BibitemOpen
  \bibfield  {author} {\bibinfo {author} {\bibfnamefont {J.~L.}\ \bibnamefont
  {Silverberg}}, \bibinfo {author} {\bibfnamefont {A.~A.}\ \bibnamefont
  {Evans}}, \bibinfo {author} {\bibfnamefont {L.}~\bibnamefont {McLeod}},
  \bibinfo {author} {\bibfnamefont {R.~C.}\ \bibnamefont {Hayward}}, \bibinfo
  {author} {\bibfnamefont {T.}~\bibnamefont {Hull}}, \bibinfo {author}
  {\bibfnamefont {C.~D.}\ \bibnamefont {Santangelo}},\ and\ \bibinfo {author}
  {\bibfnamefont {I.}~\bibnamefont {Cohen}},\ }\bibfield  {title}
  {{\selectlanguage {english}\bibinfo {title} {Using origami design principles
  to fold reprogrammable mechanical metamaterials}},\ }\href
  {https://doi.org/10.1126/science.1252876} {\bibfield  {journal} {\bibinfo
  {journal} {Science}\ }\textbf {\bibinfo {volume} {345}},\ \bibinfo {pages}
  {647–650} (\bibinfo {year} {2014})}\BibitemShut {NoStop}%
\bibitem [{\citenamefont {Sano}\ and\ \citenamefont {Wada}(2018)}]{sano2018}%
  \BibitemOpen
  \bibfield  {author} {\bibinfo {author} {\bibfnamefont {T.~G.}\ \bibnamefont
  {Sano}}\ and\ \bibinfo {author} {\bibfnamefont {H.}~\bibnamefont {Wada}},\
  }\bibfield  {title} {\bibinfo {title} {Snap-buckling in asymmetrically
  constrained elastic strips},\ }\href
  {https://doi.org/10.1103/PhysRevE.97.013002} {\bibfield  {journal} {\bibinfo
  {journal} {Physical Review E}\ }\textbf {\bibinfo {volume} {97}},\ \bibinfo
  {pages} {013002} (\bibinfo {year} {2018})}\BibitemShut {NoStop}%
\bibitem [{\citenamefont {Chaikin}\ and\ \citenamefont
  {Lubensky}(1995)}]{chaikin1995}%
  \BibitemOpen
  \bibfield  {author} {\bibinfo {author} {\bibfnamefont {P.~M.}\ \bibnamefont
  {Chaikin}}\ and\ \bibinfo {author} {\bibfnamefont {T.~C.}\ \bibnamefont
  {Lubensky}},\ }\href@noop {} {\emph {\bibinfo {title} {Principles of
  condensed matter physics}}}\ (\bibinfo  {publisher} {Cambridge University
  Press},\ \bibinfo {year} {1995})\BibitemShut {NoStop}%
\bibitem [{\citenamefont {Peskin}(2018)}]{peskin2018}%
  \BibitemOpen
  \bibfield  {author} {\bibinfo {author} {\bibfnamefont {M.}~\bibnamefont
  {Peskin}},\ }\href@noop {} {\emph {\bibinfo {title} {An introduction to
  quantum field theory}}}\ (\bibinfo  {publisher} {CRC press},\ \bibinfo {year}
  {2018})\BibitemShut {NoStop}%
\bibitem [{\citenamefont {Frisch}(1996)}]{frisch1996}%
  \BibitemOpen
  \bibfield  {author} {\bibinfo {author} {\bibfnamefont {U.}~\bibnamefont
  {Frisch}},\ }\href
  {https://www.cambridge.org/highereducation/books/turbulence/FD8C5E35E5F1CA850E017461942A59AC#overview}
  {\emph {\bibinfo {title} {Turbulence: the legacy of A.N. Kolmogorov}}}\
  (\bibinfo  {publisher} {Cambridge University Press},\ \bibinfo {year}
  {1996})\BibitemShut {NoStop}%
\bibitem [{\citenamefont {Crawford}\ and\ \citenamefont
  {Knobloch}(1991)}]{crawford1991}%
  \BibitemOpen
  \bibfield  {author} {\bibinfo {author} {\bibfnamefont {J.~D.}\ \bibnamefont
  {Crawford}}\ and\ \bibinfo {author} {\bibfnamefont {E.}~\bibnamefont
  {Knobloch}},\ }\bibfield  {title} {\bibinfo {title} {Symmetry and
  symmetry-breaking bifurcations in fluid dynamics},\ }\href@noop {} {\bibfield
   {journal} {\bibinfo  {journal} {Annual Review of Fluid Mechanics}\ }\textbf
  {\bibinfo {volume} {23}},\ \bibinfo {pages} {341–387} (\bibinfo {year}
  {1991})}\BibitemShut {NoStop}%
\bibitem [{\citenamefont {Michelin}\ and\ \citenamefont
  {Lauga}(2010)}]{michelin2010}%
  \BibitemOpen
  \bibfield  {author} {\bibinfo {author} {\bibfnamefont {S.}~\bibnamefont
  {Michelin}}\ and\ \bibinfo {author} {\bibfnamefont {E.}~\bibnamefont
  {Lauga}},\ }\bibfield  {title} {\bibinfo {title} {Efficiency optimization and
  symmetry-breaking in a model of ciliary locomotion},\ }\href
  {https://doi.org/10.1063/1.3507951} {\bibfield  {journal} {\bibinfo
  {journal} {Physics of Fluids}\ }\textbf {\bibinfo {volume} {22}},\ \bibinfo
  {pages} {111901} (\bibinfo {year} {2010})}\BibitemShut {NoStop}%
\bibitem [{\citenamefont {Tjhung}\ \emph {et~al.}(2012)\citenamefont {Tjhung},
  \citenamefont {Marenduzzo},\ and\ \citenamefont {Cates}}]{tjhung2012}%
  \BibitemOpen
  \bibfield  {author} {\bibinfo {author} {\bibfnamefont {E.}~\bibnamefont
  {Tjhung}}, \bibinfo {author} {\bibfnamefont {D.}~\bibnamefont {Marenduzzo}},\
  and\ \bibinfo {author} {\bibfnamefont {M.~E.}\ \bibnamefont {Cates}},\
  }\bibfield  {title} {\bibinfo {title} {Spontaneous symmetry breaking in
  active droplets provides a generic route to motility},\ }\href
  {https://doi.org/10.1073/pnas.1200843109} {\bibfield  {journal} {\bibinfo
  {journal} {Proceedings of the National Academy of Sciences}\ }\textbf
  {\bibinfo {volume} {109}},\ \bibinfo {pages} {12381–12386} (\bibinfo {year}
  {2012})}\BibitemShut {NoStop}%
\bibitem [{\citenamefont {Joulin}\ and\ \citenamefont
  {Vidal}(1998)}]{joulin1998}%
  \BibitemOpen
  \bibfield  {author} {\bibinfo {author} {\bibfnamefont {G.}~\bibnamefont
  {Joulin}}\ and\ \bibinfo {author} {\bibfnamefont {P.}~\bibnamefont {Vidal}},\
  }\bibfield  {title} {\bibinfo {title} {Flames, shocks and detonation},\
  }\href@noop {} {\bibfield  {journal} {\bibinfo  {journal} {Hydrodynamics and
  Nonlinear Instabilities (ed. C. Godreche \& P. Manneville)}\ ,\ \bibinfo
  {pages} {546–568}} (\bibinfo {year} {1998})}\BibitemShut {NoStop}%
\bibitem [{\citenamefont {Strogatz}(1994)}]{strogatz1994}%
  \BibitemOpen
  \bibfield  {author} {\bibinfo {author} {\bibfnamefont {S.~H.}\ \bibnamefont
  {Strogatz}},\ }\href@noop {} {\emph {\bibinfo {title} {Nonlinear dynamics and
  Chaos: with applications to physics, biology, chemistry, and engineering}}},\
  Studies in nonlinearity\ (\bibinfo  {publisher} {Addison-Wesley Pub},\
  \bibinfo {year} {1994})\BibitemShut {NoStop}%
\bibitem [{\citenamefont {Cosserat}\ and\ \citenamefont
  {Cosserat}(1909)}]{cosserat1909}%
  \BibitemOpen
  \bibfield  {author} {\bibinfo {author} {\bibfnamefont {E.}~\bibnamefont
  {Cosserat}}\ and\ \bibinfo {author} {\bibfnamefont {F.}~\bibnamefont
  {Cosserat}},\ }\href@noop {} {\emph {\bibinfo {title} {Th{\'e}orie des corps
  d{\'e}formables}}}\ (\bibinfo  {publisher} {A. Hermann et fils},\ \bibinfo
  {year} {1909})\BibitemShut {NoStop}%
\bibitem [{\citenamefont {Gazzola}\ \emph {et~al.}(2018)\citenamefont
  {Gazzola}, \citenamefont {Dudte}, \citenamefont {McCormick},\ and\
  \citenamefont {Mahadevan}}]{gazzola2018}%
  \BibitemOpen
  \bibfield  {author} {\bibinfo {author} {\bibfnamefont {M.}~\bibnamefont
  {Gazzola}}, \bibinfo {author} {\bibfnamefont {L.~H.}\ \bibnamefont {Dudte}},
  \bibinfo {author} {\bibfnamefont {A.~G.}\ \bibnamefont {McCormick}},\ and\
  \bibinfo {author} {\bibfnamefont {L.}~\bibnamefont {Mahadevan}},\ }\bibfield
  {title} {\bibinfo {title} {Forward and inverse problems in the mechanics of
  soft filaments},\ }\href {https://doi.org/10.1098/rsos.171628} {\bibfield
  {journal} {\bibinfo  {journal} {Royal Society Open Science}\ }\textbf
  {\bibinfo {volume} {5}},\ \bibinfo {pages} {171628} (\bibinfo {year}
  {2018})}\BibitemShut {NoStop}%
\bibitem [{\citenamefont {Radisson}\ and\ \citenamefont
  {Kanso}()}]{radisson2022PRE}%
  \BibitemOpen
  \bibfield  {author} {\bibinfo {author} {\bibfnamefont {B.}~\bibnamefont
  {Radisson}}\ and\ \bibinfo {author} {\bibfnamefont {E.}~\bibnamefont
  {Kanso}},\ }\bibfield  {title} {\bibinfo {title} {Dynamic behavior of an
  elastic strip in the vicinity of a shape transition},\ }\href@noop {}
  {\bibinfo  {journal} {submitted to Physical Review E}\ }\BibitemShut
  {NoStop}%
\bibitem [{\citenamefont {Reid}\ \emph {et~al.}(2017)\citenamefont {Reid},
  \citenamefont {Lechenault}, \citenamefont {Rica},\ and\ \citenamefont
  {Adda-Bedia}}]{reid2017}%
  \BibitemOpen
\bibfield  {journal} {  }\bibfield  {author} {\bibinfo {author} {\bibfnamefont
  {A.}~\bibnamefont {Reid}}, \bibinfo {author} {\bibfnamefont {F.}~\bibnamefont
  {Lechenault}}, \bibinfo {author} {\bibfnamefont {S.}~\bibnamefont {Rica}},\
  and\ \bibinfo {author} {\bibfnamefont {M.}~\bibnamefont {Adda-Bedia}},\
  }\bibfield  {title} {{\selectlanguage {english}\bibinfo {title} {Geometry and
  design of origami bellows with tunable response}},\ }\href
  {https://doi.org/10.1103/PhysRevE.95.013002} {\bibfield  {journal} {\bibinfo
  {journal} {Physical Review E}\ }\textbf {\bibinfo {volume} {95}},\ \bibinfo
  {pages} {013002} (\bibinfo {year} {2017})}\BibitemShut {NoStop}%
\end{thebibliography}%


%apsrev4-2.bst 2019-01-14 (MD) hand-edited version of apsrev4-1.bst
%Control: key (0)
%Control: author (8) initials jnrlst
%Control: editor formatted (1) identically to author
%Control: production of article title (0) allowed
%Control: page (0) single
%Control: year (1) truncated
%Control: production of eprint (0) enabled
\begin{thebibliography}{10}%
\makeatletter
\providecommand \@ifxundefined [1]{%
 \@ifx{#1\undefined}
}%
\providecommand \@ifnum [1]{%
 \ifnum #1\expandafter \@firstoftwo
 \else \expandafter \@secondoftwo
 \fi
}%
\providecommand \@ifx [1]{%
 \ifx #1\expandafter \@firstoftwo
 \else \expandafter \@secondoftwo
 \fi
}%
\providecommand \natexlab [1]{#1}%
\providecommand \enquote  [1]{``#1''}%
\providecommand \bibnamefont  [1]{#1}%
\providecommand \bibfnamefont [1]{#1}%
\providecommand \citenamefont [1]{#1}%
\providecommand \href@noop [0]{\@secondoftwo}%
\providecommand \href [0]{\begingroup \@sanitize@url \@href}%
\providecommand \@href[1]{\@@startlink{#1}\@@href}%
\providecommand \@@href[1]{\endgroup#1\@@endlink}%
\providecommand \@sanitize@url [0]{\catcode `\\12\catcode `\$12\catcode
  `\&12\catcode `\#12\catcode `\^12\catcode `\_12\catcode `\%12\relax}%
\providecommand \@@startlink[1]{}%
\providecommand \@@endlink[0]{}%
\providecommand \url  [0]{\begingroup\@sanitize@url \@url }%
\providecommand \@url [1]{\endgroup\@href {#1}{\urlprefix }}%
\providecommand \urlprefix  [0]{URL }%
\providecommand \Eprint [0]{\href }%
\providecommand \doibase [0]{https://doi.org/}%
\providecommand \selectlanguage [0]{\@gobble}%
\providecommand \bibinfo  [0]{\@secondoftwo}%
\providecommand \bibfield  [0]{\@secondoftwo}%
\providecommand \translation [1]{[#1]}%
\providecommand \BibitemOpen [0]{}%
\providecommand \bibitemStop [0]{}%
\providecommand \bibitemNoStop [0]{.\EOS\space}%
\providecommand \EOS [0]{\spacefactor3000\relax}%
\providecommand \BibitemShut  [1]{\csname bibitem#1\endcsname}%
\let\auto@bib@innerbib\@empty
%</preamble>
\bibitem [{\citenamefont {Strogatz}(1994)}]{strogatz1994}%
  \BibitemOpen
  \bibfield  {author} {\bibinfo {author} {\bibfnamefont {S.~H.}\ \bibnamefont
  {Strogatz}},\ }\href@noop {} {\emph {\bibinfo {title} {Nonlinear dynamics and
  Chaos: with applications to physics, biology, chemistry, and engineering}}},\
  Studies in nonlinearity\ (\bibinfo  {publisher} {Addison-Wesley Pub},\
  \bibinfo {year} {1994})\BibitemShut {NoStop}%
\bibitem [{\citenamefont {Gomez}\ \emph {et~al.}(2017)\citenamefont {Gomez},
  \citenamefont {Moulton},\ and\ \citenamefont {Vella}}]{gomez2017}%
  \BibitemOpen
  \bibfield  {author} {\bibinfo {author} {\bibfnamefont {M.}~\bibnamefont
  {Gomez}}, \bibinfo {author} {\bibfnamefont {D.}~\bibnamefont {Moulton}},\
  and\ \bibinfo {author} {\bibfnamefont {D.}~\bibnamefont {Vella}},\ }\bibfield
   {title} {\bibinfo {title} {Critical slowing down in purely elastic
  ‘snap-through’ instabilities},\ }\href
  {https://doi.org/10.1038/nphys3915} {\bibfield  {journal} {\bibinfo
  {journal} {Nature Physics}\ }\textbf {\bibinfo {volume} {13}},\ \bibinfo
  {pages} {142–145} (\bibinfo {year} {2017})}\BibitemShut {NoStop}%
\bibitem [{\citenamefont {Gazzola}\ \emph {et~al.}(2018)\citenamefont
  {Gazzola}, \citenamefont {Dudte}, \citenamefont {McCormick},\ and\
  \citenamefont {Mahadevan}}]{gazzola2018}%
  \BibitemOpen
  \bibfield  {author} {\bibinfo {author} {\bibfnamefont {M.}~\bibnamefont
  {Gazzola}}, \bibinfo {author} {\bibfnamefont {L.~H.}\ \bibnamefont {Dudte}},
  \bibinfo {author} {\bibfnamefont {A.~G.}\ \bibnamefont {McCormick}},\ and\
  \bibinfo {author} {\bibfnamefont {L.}~\bibnamefont {Mahadevan}},\ }\bibfield
  {title} {\bibinfo {title} {Forward and inverse problems in the mechanics of
  soft filaments},\ }\href {https://doi.org/10.1098/rsos.171628} {\bibfield
  {journal} {\bibinfo  {journal} {Royal Society Open Science}\ }\textbf
  {\bibinfo {volume} {5}},\ \bibinfo {pages} {171628} (\bibinfo {year}
  {2018})}\BibitemShut {NoStop}%
\bibitem [{\citenamefont {Sano}\ and\ \citenamefont {Wada}(2018)}]{sano2018}%
  \BibitemOpen
  \bibfield  {author} {\bibinfo {author} {\bibfnamefont {T.~G.}\ \bibnamefont
  {Sano}}\ and\ \bibinfo {author} {\bibfnamefont {H.}~\bibnamefont {Wada}},\
  }\bibfield  {title} {\bibinfo {title} {Snap-buckling in asymmetrically
  constrained elastic strips},\ }\href
  {https://doi.org/10.1103/PhysRevE.97.013002} {\bibfield  {journal} {\bibinfo
  {journal} {Physical Review E}\ }\textbf {\bibinfo {volume} {97}},\ \bibinfo
  {pages} {013002} (\bibinfo {year} {2018})}\BibitemShut {NoStop}%
\bibitem [{\citenamefont {Timoshenko}\ and\ \citenamefont
  {Gere}(2009)}]{timoshenko2009}%
  \BibitemOpen
  \bibfield  {author} {\bibinfo {author} {\bibfnamefont {S.~P.}\ \bibnamefont
  {Timoshenko}}\ and\ \bibinfo {author} {\bibfnamefont {J.~M.}\ \bibnamefont
  {Gere}},\ }\href@noop {} {\emph {\bibinfo {title} {Theory of elastic
  stability}}}\ (\bibinfo  {publisher} {Courier Corporation},\ \bibinfo {year}
  {2009})\BibitemShut {NoStop}%
\bibitem [{\citenamefont {Pandey}\ \emph {et~al.}(2014)\citenamefont {Pandey},
  \citenamefont {Moulton}, \citenamefont {Vella},\ and\ \citenamefont
  {Holmes}}]{pandey2014}%
  \BibitemOpen
  \bibfield  {author} {\bibinfo {author} {\bibfnamefont {A.}~\bibnamefont
  {Pandey}}, \bibinfo {author} {\bibfnamefont {D.~E.}\ \bibnamefont {Moulton}},
  \bibinfo {author} {\bibfnamefont {D.}~\bibnamefont {Vella}},\ and\ \bibinfo
  {author} {\bibfnamefont {D.~P.}\ \bibnamefont {Holmes}},\ }\bibfield  {title}
  {\bibinfo {title} {Dynamics of snapping beams and jumping poppers},\ }\href
  {https://doi.org/10.1209/0295-5075/105/24001} {\bibfield  {journal} {\bibinfo
   {journal} {EPL (Europhysics Letters)}\ }\textbf {\bibinfo {volume} {105}},\
  \bibinfo {pages} {24001} (\bibinfo {year} {2014})}\BibitemShut {NoStop}%
\bibitem [{\citenamefont {Nayfeh}\ and\ \citenamefont
  {Emam}(2008)}]{nayfeh2008}%
  \BibitemOpen
  \bibfield  {author} {\bibinfo {author} {\bibfnamefont {A.~H.}\ \bibnamefont
  {Nayfeh}}\ and\ \bibinfo {author} {\bibfnamefont {S.~A.}\ \bibnamefont
  {Emam}},\ }\bibfield  {title} {\bibinfo {title} {Exact solution and stability
  of postbuckling configurations of beams},\ }\href
  {https://doi.org/10.1007/s11071-008-9338-2} {\bibfield  {journal} {\bibinfo
  {journal} {Nonlinear Dynamics}\ }\textbf {\bibinfo {volume} {54}},\ \bibinfo
  {pages} {395–408} (\bibinfo {year} {2008})}\BibitemShut {NoStop}%
\bibitem [{\citenamefont {Howell}\ \emph {et~al.}(2009)\citenamefont {Howell},
  \citenamefont {Kozyreff},\ and\ \citenamefont {Ockendon}}]{howell2009}%
  \BibitemOpen
  \bibfield  {author} {\bibinfo {author} {\bibfnamefont {P.}~\bibnamefont
  {Howell}}, \bibinfo {author} {\bibfnamefont {G.}~\bibnamefont {Kozyreff}},\
  and\ \bibinfo {author} {\bibfnamefont {J.}~\bibnamefont {Ockendon}},\
  }\href@noop {} {\emph {\bibinfo {title} {Applied solid mechanics}}},\
  \bibinfo {number} {43}\ (\bibinfo  {publisher} {Cambridge University Press},\
  \bibinfo {year} {2009})\BibitemShut {NoStop}%
\bibitem [{\citenamefont {Gomez}(2018)}]{gomez2018}%
  \BibitemOpen
  \bibfield  {author} {\bibinfo {author} {\bibfnamefont {M.}~\bibnamefont
  {Gomez}},\ }\emph {\bibinfo {title} {Ghosts and bottlenecks in elastic
  snap-through}},\ \href@noop {} {Ph.D. thesis},\ \bibinfo  {school}
  {University of Oxford / University of Oxford} (\bibinfo {year}
  {2018})\BibitemShut {NoStop}%
\bibitem [{\citenamefont {Radisson}\ and\ \citenamefont
  {Kanso}()}]{radisson2022PRE}%
  \BibitemOpen
  \bibfield  {author} {\bibinfo {author} {\bibfnamefont {B.}~\bibnamefont
  {Radisson}}\ and\ \bibinfo {author} {\bibfnamefont {E.}~\bibnamefont
  {Kanso}},\ }\bibfield  {title} {\bibinfo {title} {Dynamic behavior of elastic
  strips near shape transition},\ }\href@noop {} {\bibinfo  {journal}
  {submitted to Physical Review E}\ }\BibitemShut {NoStop}%
\end{thebibliography}%

\end{document}

\title{SUPPLEMENTAL DOCUMENT\\[2.5ex] Elastic snap-through instabilities are governed by geometric symmetries}
	\author{Basile Radisson, Eva Kanso}
	\affiliation{Department of Aerospace and Mechanical Engineering, University of Southern California, Los Angeles, CA 90089-1191, USA}
	\date{\today}

	\maketitle
	
\section{Symmetry-breaking of a pitchfork bifurcation: a canonical example}\label{sec:symmetryPitchfork}
We review the canonical example of symmetry-breaking in a system with pitchfork bifurcation. It is well-known that starting from a system with pitchfork bifurcation symmetry, the introduction of an additional parameter that breaks the left-right symmetry could turn the pitchfork bifurcation into a saddle-node-like bifurcation. 

\bigskip
\par
\begin{figure}[t]
	\centering
	\includegraphics[width =\textwidth]{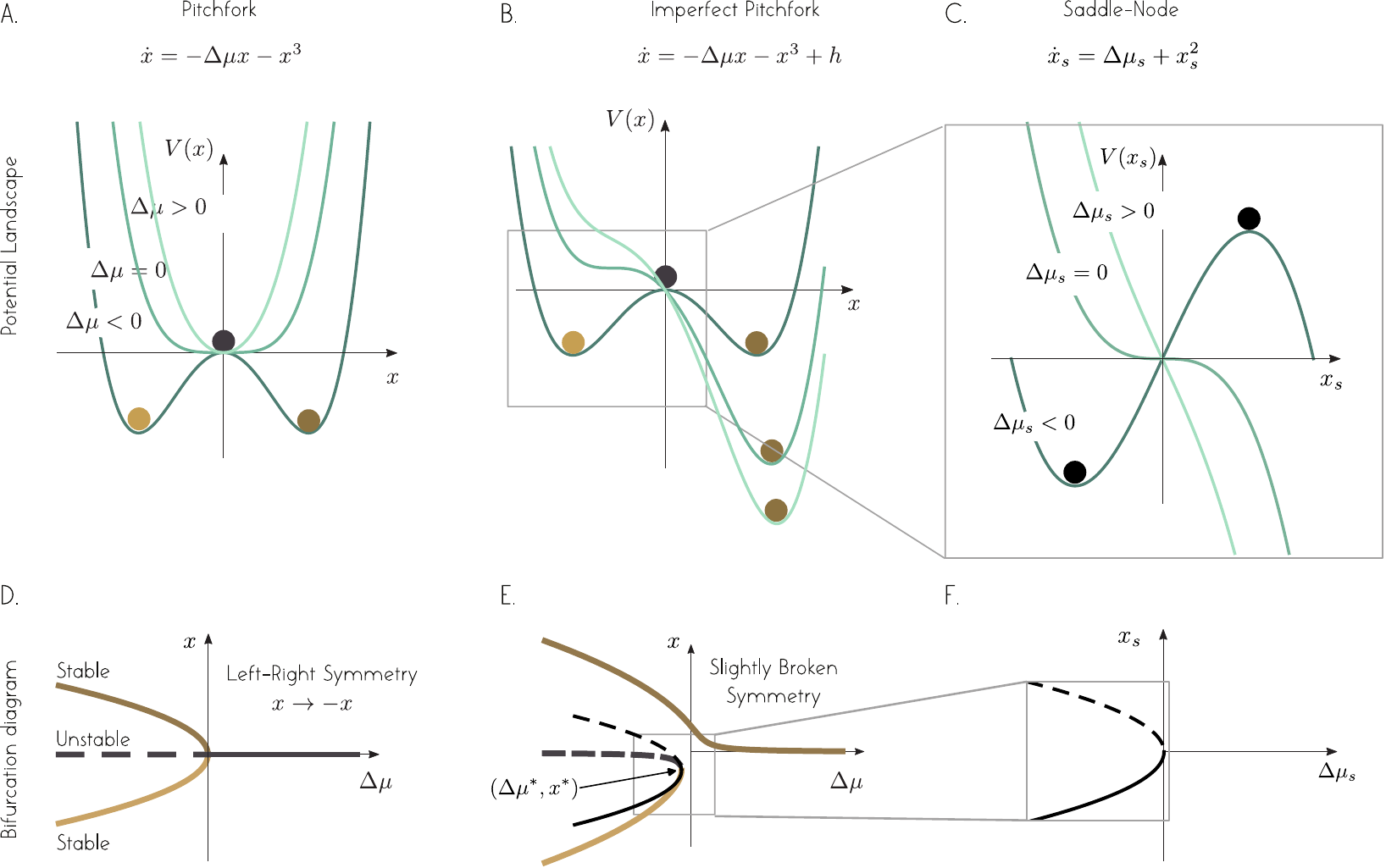}
	\caption{{\textbf{Pitchfork bifurcation and symmetries}  Pitchfork bifurcations are known to arise in systems with binary symmetries (i.e left-right, top-bottom, etc). Breaking this symmetry alters the nature of the transition into a saddle-node bifurcation.  (First row) Potential landscape (green lines) associated with a  \textbf{A.} supercritical pitchfork,  \textbf{B.} imperfect supercritical pitchfork, \textbf{C.} saddle-node bifurcation for different values of the bifurcation parameter $\Delta\mu$. Equilibria (dot symbols) are highlighted for each system. (Second row)  Bifurcation diagrams representing the evolution of these equilibria as a function of the bifurcation parameter (full lines for stable equilibria and dashed lines for unstable) for \textbf{D.} the system with supercritical pitchfork, \textbf{E.} the system with broken symmetry exhibiting an imperfect pitchfork bifurcation, and \textbf{F.} the system with saddle-node bifurcation. Note the local equivalence (see grey boxes in panels B and E) between the imperfect pitchfork and saddle-node bifurcations.}}
	\label{fig:symmetryPitchfork}
\end{figure}

\noindent
\textbf{Canonical saddle node bifurcation.} The canonical form of a saddle-node bifurcation for a one-degree-of-freedom system (say $x_s$) is given by
%-----
\begin{equation}
    \dot{x}_s=\Delta\mu_s +x_s^2.
    \label{eq:saddleNode1stOrder}
\end{equation}
%-----------
Here, the dot stands for the first order derivative with respect to time and $\Delta \mu_s$ is the bifurcation parameter. This system admits a potential function $V(x_s)$ such that $\dot{x}_s=-dV(x_s)/dx_s$,
%-------
\begin{equation}
    V(x_s)=-\Delta\mu x_s-\frac{1}{3}x_s^3.
    \label{eq:saddleNode1stOrderPotential}
\end{equation}
%--------
For $\Delta \mu < 0$, \eqref{eq:saddleNode1stOrderPotential} admits  one potential well and one energy barrier (Fig. \ref{fig:symmetryPitchfork}C), reflecting one stable equilibrium and one unstable equilibrium of~\eqref{eq:saddleNode1stOrder}. The two equilibria merge and disappear in $\Delta \mu = 0$, as pictured on the bifurcation diagram (Fig. \ref{fig:symmetryPitchfork}F).

\bigskip
\par\noindent
\textbf{Canonical pitchfork bifurcation.} The canonical form of a supercritical pitchfork bifurcation is given by
\begin{equation}
    \dot{x}=-\Delta\mu x-x^3.
    \label{eq:pitchfork1stOrder}
\end{equation}
This system admits a potential function,
\begin{equation}
    V(x)=\frac{1}{2}\Delta\mu x^2+\frac{1}{4}x^4.
    \label{eq:pitchfork1stOrderPotential}
\end{equation}
Pitchfork bifurcations possess a left-right symmetry. Here, this is reflected by the invariance of \eqref{eq:pitchfork1stOrder} and \eqref{eq:pitchfork1stOrderPotential} through a transformation $x\rightarrow-x$. 

For $\Delta\mu<0$, the potential $V(x)$ exhibits two potential wells that are symmetrically distributed around the energy barrier at the origin $x=0$ (Fig. S.1A). As $\Delta\mu$ increases, the potential landscape gets deformed symmetrically until, at $\Delta\mu=0$, the two potential wells merge with the energy barrier at the origin. For $\Delta\mu>0$, only a single potential well exists at the origin.
%see Fig. \ref{fig:fig:symmetryPitchfork}D where the left-right symmetry is visible.

\bigskip
\par\noindent
\textbf{Canonical symmetry-breaking of a pitchfork bifurcation.} 
When an asymmetry is introduced in~\eqref{eq:pitchfork1stOrder}, 
it turns it into an imperfect pitchfork,
\begin{equation}
    \dot{x}=-\Delta\mu x-x^3+h,
    \label{eq:imperfectPitchfork}
\end{equation} 
for which the potential function is given by
\begin{equation}
    V(x)=\frac{1}{2}\Delta\mu x^2+\frac{1}{4}x^4-hx.
\end{equation}
Here, the parameter $h>0$ represents the asymmetry. This changes the nature of the bifurcation dramatically. Even for $|h|\ll  1$, the loss of symmetry makes one of the potential wells reach the energy barrier before the other, leading to a sudden disappearance of these two equilibria -- the stable equilibrium at the potential well and the unstable equilibrium at the energy barrier (Fig. \ref{fig:symmetryPitchfork}B). Locally, at the bifurcation point where these two equilibria merge
%unstable equilibrium merges with one of the two stable ones 
before disappearing, the bifurcation has the form of a saddle-node \cite{strogatz1994} (see Fig. \ref{fig:symmetryPitchfork}C and \ref{fig:symmetryPitchfork}F). 

This is formally demonstrated as follows. We remark that this bifurcation occurs when the local minima of the RHS $f(x, \Delta\mu)=-\Delta\mu x-x^3+h$ in \eqref{eq:imperfectPitchfork} hits the zero axis. We solve for $x^*$ such that $f'(x^*, \Delta\mu)=0$ and $\Delta\mu^*$ such that $f(x^*, \Delta\mu^*)=0$ and find the coordinates of the bifurcation point (see Fig. \ref{fig:symmetryPitchfork}E):

\begin{equation}
    \Delta\mu^{*}=-3\left(\frac{h}{2}\right)^{2/3},\qquad x^*=-\sqrt{-\frac{\Delta\mu^{*}}{3}}.
\end{equation}

We then seek for an asymptotic approximation of \eqref{eq:imperfectPitchfork} in the vicinity of this bifurcation point. We set $x_s=x-x^*$ and $\Delta\mu_s=\Delta\mu-\Delta\mu^*$ so that the bifurcation occurs in $(\Delta\mu_s=0, x_s=0)$. We introduce these new variables in \eqref{eq:imperfectPitchfork} and write an asymptotic expansion in the limit $x_s \ll 1$, $\Delta\mu_s \ll 1$. We get:

\begin{equation}
    \dot{x}_s=|x^*|(\Delta\mu_s+3x_s^2)+O(\Delta\mu^{3/2}),
    \label{eq:saddleImperfect}
\end{equation}

which is precisely (up to numerical constants) the normal form of a saddle-node (see \eqref{eq:saddleNode1stOrder}). The bifurcation diagram associated with this asymptotic approximation (black lines) is compared to the bifurcation diagram of the imperfect pitchfork in Fig. \ref{fig:symmetryPitchfork}E and compares well with it in the very vicinity of the bifurcation point.

We note that the symmetry breaking mechanism, reviewed here, that turns a pitchfork into a saddle-node bifurcation also stands for a subcritical pitchfork bifurcation, whose normal form is simply obtained by applying a transformation $(x\rightarrow-x,\ t\rightarrow-t)$ to \eqref{eq:pitchfork1stOrder}.

%%%%%%%%%%%%%%%%%%%%%%

\begin{table}[!b]
\caption{Strip's geometric and material parameters}
\begin{equation*}
\begin{array}{|c | c | c|}	
\hline
\multicolumn{3}{|c|}{\textbf{Numerics}} \\
\hline
\textbf{Parameter} & \textbf{Value} & \textbf{units}\\
\hline
L & 0.156	& \text{meters (m)}\\
\hline
b & 2\times 10^{-2}	& \text{meters (m)}\\
\hline
h & 1\times 10^{-3}	& \text{meters (m)}\\
\hline
\rho & 1.35\times 10^{3}	& \text{kilograms per meter cube (Kg$\cdot$m$^{-3}$)}\\
\hline
E & 3.7\times 10^{9}	& \text{Pascals (Pa)}\\
\hline
G & 3.7\times 10^{9}	& \text{Pascals (Pa) }\\
\hline
S_1=S_2=S_3 & 1\times 10^{5}	& \text{Newtons (N)}\\
\hline
\end{array} \qquad 
\begin{array}{|l|}	
\hline
\multicolumn{1}{|c|}{\textbf{Experiments of Gomez et al.~\cite{gomez2017}}} \\
\hline
%\textbf{Parameter} & \textbf{Value} & \textbf{units}\\
%\hline
\multicolumn{1}{|c|}{\text{Polyethylene terephthalate (PET)}} \\
\hline
\rho = 1.337\times 10^{3} \quad	 \text{Kg$\cdot$m$^{-3}$}\\
%\hline
E = 5.707\times 10^{9}	\quad \text{Pa}\\
L \in \{0.240, 0.290, 0.430\} \ \text{m} \\
\hline
\multicolumn{1}{|c|}{\text{Stainless steel}} \\
\hline
\rho = 7.881\times 10^{3}\quad	 \text{Kg$\cdot$m$^{-3}$}\\
%\hline
E = 203.8\times 10^{9}	\quad \text{Pa}\\
L \in \{0.140, 0.280\} \ \text{m} \\
\hline
\end{array}
\label{eq:cosseratRodParameters}
\end{equation*}
\label{tab:param}
\end{table}

%%%%%%%%%%%%%%%%%
\section{Mathematical methods}
\label{sec:math}

%\subsection{3D Cosserat Rod Theory}
%\label{sec:cosserat}
\par\noindent
\textbf{A. Numerical simulations based on the Cosserat rod theory.} We consider an elastic strip of length $L$ and rectangular cross-section of width $b$ and thickness $h$; see, e.g., Fig.~\ref{fig:towardInitialCondition}. We numerically integrate the discrete Cosserat rod equations governing the dynamics of the strip's centerline $\mathbf{r}(s,t)$ and orientation tensor $\mathbf{Q}(s,t)$ using our own implementation of the method described in \cite{gazzola2018}.
%encoding the orientation of the strip's directors, i.e., unit vectors $\{\mathbf{d}_1,\mathbf{d}_2,\mathbf{d}_3\}$ where ($\mathbf{d}_1,\mathbf{d}_2$) spans the strip's cross-sectional area. 
Here, $s$ is arclength and $t$ is time. We integrate the discrete equations forward in time to obtain the equilibrium configuration of the Euler-buckled strip, as well as the equilibrium configuration of the buckled elastic strip under further translational and rotational boundary actuation.
In all numerical simulations, we use the set of dimensional parameters listed in Table~\ref{tab:param}.  This set of parameters corresponds physically to strips made from plastic sheets as in the experiments of \cite{sano2018}. Throughout the study, %we set $S_1$, $S_2$, $S_3$ to the large value in Table~\ref{tab:param} to limit the stretchability of the strip and ensure that the inextensibility condition is weakly enforced. 
we set  $\Delta L=L/100$, except for the snapshots showing the 3D rendering where we used $\Delta L=L/20$ for illustration purposes. 
%The dissipation parameter is set to $\nu=10^{-4}kg\cdot(ms)^{-1}$ for the numerical data plotted in \textcolor{red}{Fig. 2, 4B, \ref{fig:linearstability}, \ref{fig:stabilityNumerics}, \ref{fig:fig5}A, \ref{fig:fig5}B, \ref{fig:fig5}C and to $\nu=3 \times 10^{-4}kg\cdot(ms)^{-1}$ for the numerical data plotted in Fig. 4A, 4C, \ref{fig:taperedBeam}}.

%\subsection{Euler Beam Model}
%\label{sec:euler}
\bigskip
\par\noindent
\textbf{B. Euler beam model.}
We also analyze the equilibria of the elastic strip and their stability using the Euler beam model, which affords semi-analytical results. In the small deflection limit, the non-dimensional transverse displacement $W$ of the elastic strip in the $y$-direction is described by the linear Euler beam equation \cite{timoshenko2009}, which in  non dimensional form is given by
%------
\begin{equation}
\frac{\partial^2 W}{\partial T^2}+\frac{\partial^4W}{\partial X ^4}+\Lambda^2 \frac{\partial^2 W}{\partial X^2}=0.
\label{eq:beam_equation_nodim}
\end{equation}
%----
Here, $\Lambda^2$ is the non-dimensional longitudinal compression applied to the elastic strip, $X\in[-1/2,1/2]$ is the non dimensional longitudinal coordinate and $T$ the non-dimensional time.

This partial differential equation  is complemented by the non-linear geometric (incompressibility) constraint \cite{pandey2014, gomez2017},
%-------
\begin{equation}
\int_{-1/2}^{1/2}\left(\frac{\partial W}{\partial X}\right)^2dX=2 .
\label{eq:geometrical_constraint_nodim}
\end{equation}
%-------
%that expresses the condition for the shape of the strip to satisfy the end to end shortening applied by the boundaries \cite{pandey2014, gomez2017}.
The system formed by \eqref{eq:beam_equation_nodim} and \eqref{eq:geometrical_constraint_nodim} must be complemented by a set of four appropriate boundary conditions. These boundary conditions depend on the different type of boundary actuation as summarized in table \ref{tab:dimensional_boundary_conditions}.
%%%%%%%%%%%%%%%%%%%%%%%%%%%%

\bigskip
\par\noindent
\textbf{C. Static equilibria.}
The static equilibria $W_\textrm{eq}(X)$ of the elastic strip are solutions of the steady counterpart of~\eqref{eq:beam_equation_nodim},
%------
\begin{equation}
\frac{d^4 W_\textrm{eq}}{d X ^4}+\Lambda_\textrm{eq}^2 \frac{d^2 W_\textrm{eq}}{d X^2}=0,
\label{eq:beam_equation_nodim_static}
\end{equation}
%-------
whose general solution is of the form
%-----
\begin{equation}
W_\textrm{eq}(X)=A\sin(\Lambda_\textrm{eq} X)+B\cos(\Lambda_\textrm{eq} X)+ CX+D.
\label{eq:sol_general_static}
\end{equation}
%------
Here, $A$, $B$, $C$, $D$ are 4 unknown constants that must be chosen so that \eqref{eq:sol_general_static} satisfies the appropriate boundary conditions given in table \ref{tab:dimensional_boundary_conditions}. %\eqref{eq:adim_misalignment} or
%\eqref{eq:boundary_smallslope_adim}. 
%Writing the CC, CH, or HH 
Writing the boundary conditions of the elastic strip yields a system of equations of the form,
%-------
\begin{equation}
	\mathbf{M}\mathbf{v}=\mathbf{b},
	\label{eq:linearSystem}
\end{equation}
%--------
where $\mathbf{v}=(A,B,C,D)$.  The geometric constraint
 \eqref{eq:geometrical_constraint_nodim} implies that the equilibrium configurations must also satisfy 
 %---
 \begin{equation}
 \int_{-1/2}^{1/2} \left(\dfrac{\partial W_\textrm{eq}}{\partial X}\right)^2 dX = 2.
 \label{eq:geometrical_constraint_equilibrium}
 \end{equation}
 %---
The system of equations in \eqref{eq:linearSystem} and \eqref{eq:geometrical_constraint_equilibrium} determines the eigenvalue $\Lambda_{\textrm{eq}}$ and eigenfunction $W_\textrm{eq}(X)$ by providing conditions to solve for $\Lambda_\textrm{eq}$ and $(A,B,C,D)$, as discussed in detail in sections~\ref{sec:buckling}, \ref{sec:translation}, and~\ref{sec:rotation}.

%\subsection{Linear Stability Analysis}

\section{Euler Buckling}
\label{sec:buckling}

An initially straight elastic strip subject to a compression force undergoes a buckling instability through a supercritical pitchfork bifurcation beyond a given threshold (see, e.g.,~\cite{nayfeh2008, howell2009}). The threshold at which this instability occurs depends on the boundary conditions. Four types of boundary conditions are typically discussed in classic texts (see, e.g.,~\cite{timoshenko2009}): clamped-clamped (CC), hinged-hinged (HH), clamped-hinged (CH), and clamped-free. The clamped-free case has the lowest threshold, to this buckling instability and the clamped-clamped the highest. Here, we are interested in the first three boundary conditions: CC, HH, and CH. 

\bigskip
\par
\begin{figure}[!t]
	\centering
	\includegraphics[width =\textwidth]{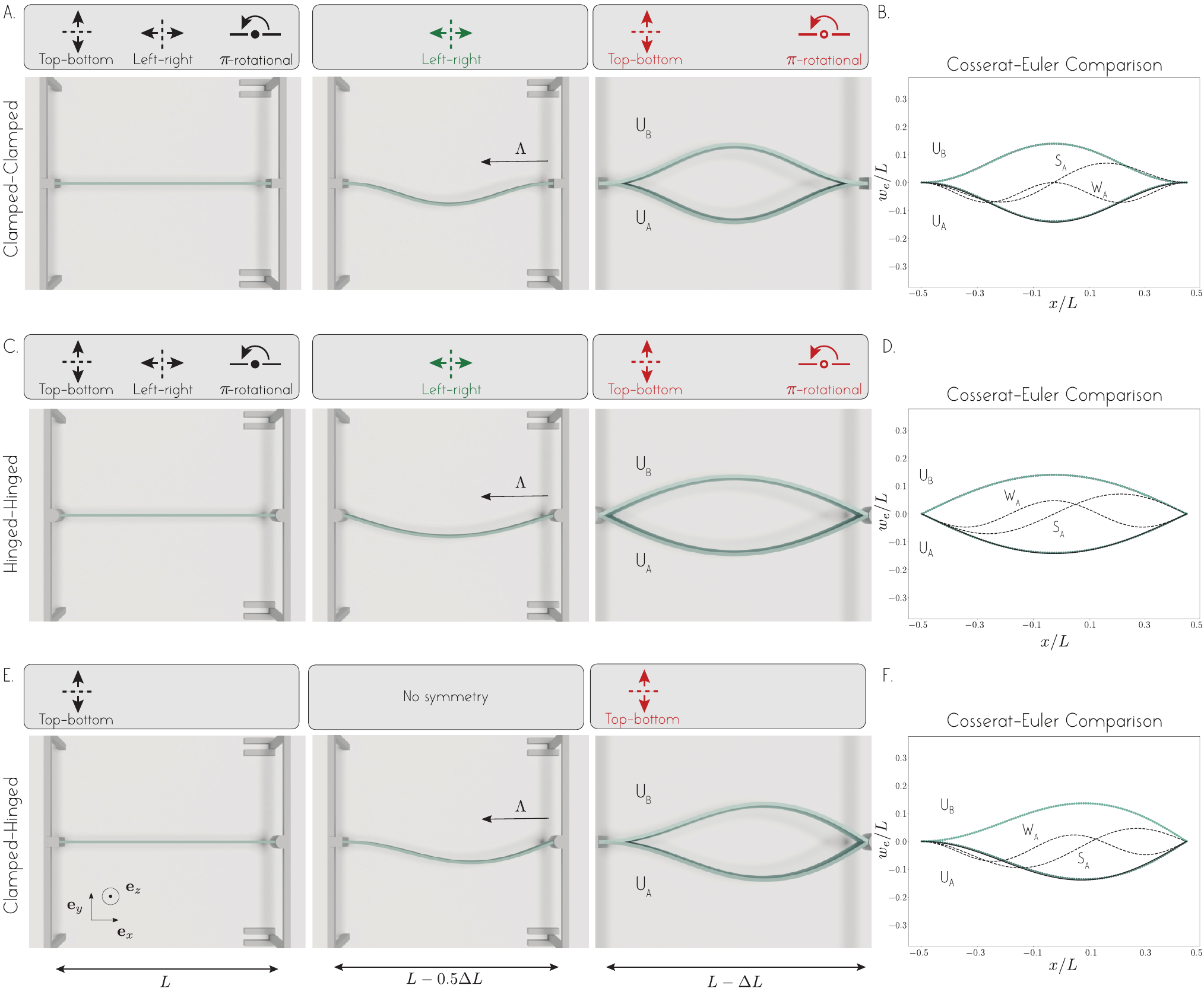}
	\caption{{\textbf{Euler buckling and symmetries} When the two ends of a strip are pushed towards each other, the strip passes from a straight configuration (left column of \textbf{A,C,E}) to a buckled configuration following a standard Euler-buckling instability. This is true for clamped (\textbf{A,B}), hinged (\textbf{C,D}), or mixed boundary conditions (\textbf{E,F}). In each case, the system is bistable, with two stable buckled states U\textsubscript{A} and U\textsubscript{B}. The original straight beam admits different symmetries depending on the type of boundary conditions (shown in black in the top panels in the left column of \textbf{A,C,E}). When buckling occurs, the buckled configuration conserves some of these symmetries (shown in green in the middle left column of \textbf{A,C,E}) and breaks others (shown in red in the middle right column of \textbf{A,C,E}). Conserved symmetries map a solution to itself and broken symmetries map a solution to its twin solution (see Fig.~\ref{fig:eulerBucklingSymmetries}). 
 (\textbf{B, D, F}) Numerical solutions (green lines) are compared to analytical solutions based on the Euler beam model (black lines).
%While not naturally selected by the system, the S-equilibrium configuraton is obtained  by imposing a zero deflection condition on the midpoint of the centerline of the Cosserat rod.
Higher order modes of buckling S\textsubscript{A} and W\textsubscript{A} (black dashed lines) the Euler beam model are superimposed.}}
	\label{fig:towardInitialCondition}
\end{figure}

\noindent
\textbf{A. Equilibria.}
For each set of boundary conditions, the buckled strip admits an infinite hierarchy of static equilibria that come in pairs of increasing value of bending energy $\mathcal{E}_b$. The pair of static equilibria that share the smallest energy level have a U like shape, and thus we label them U\textsubscript{A} and U\textsubscript{B}. The pair of  equilibria at the next energy level has an S like shape and are labeled S\textsubscript{A} and S\textsubscript{B} and the pair of equilibria at the third energy level has a W like shape and is labeled W\textsubscript{A} and W\textsubscript{B}. These static equilibria are obtained by two methods: numerically using the Cosserat rod theory (Fig. \ref{fig:towardInitialCondition}) and semi-analytically using the Euler beam model. The semi-analytic solutions are listed in Table \ref{tab:dimensional_boundary_conditions}. Contrary to the referential used in the paper, in Table \ref{tab:dimensional_boundary_conditions}, $x$ is measured from the left end of the strip.

\bigskip
\par\noindent
\textbf{B. Stability.}
Linear stability analysis of the static equilibria of the Euler-buckled strip in the CC, HH, and CH cases shows that only the fundamental harmonic pair U\textsubscript{A,B} is stable; all subsequent, even and odd, harmonic pairs are unstable, including S\textsubscript{A,B} and W\textsubscript{A,B}.

%%%%%%%%%%%%%%%%%%%%%%%%%%%%%%%%%%%%%%%%%%%%%%%%%%%%

%\section{Invariance and Symmetries in Euler Buckling}
%\label{sec:Eulersymmetries}

\bigskip
\par\noindent
\textbf{C. Invariance and symmetries in Euler buckling.}
The equations of motion governing the elastic strip are invariant under the following three transformations: the top-down transformation $w\rightarrow-w$, the left-right transformation $x\rightarrow-x$, and the $\pi$-rotational transformation $w\rightarrow-w$ and $x\rightarrow-x$. This invariance is not unique to the Euler beam model; it is also a characteristic of the 3D Cosserat rod theory. 
The CC and HH boundary conditions respect the invariance of the governing equations under all three transformations, while the CH boundary conditions respect only invariance under the top-down transformation; see Fig.~\ref{fig:towardInitialCondition} and Table~\ref{tab:dimensional_boundary_conditions}.

\begin{figure}[!t]
	\centering
	\includegraphics[width =\textwidth]{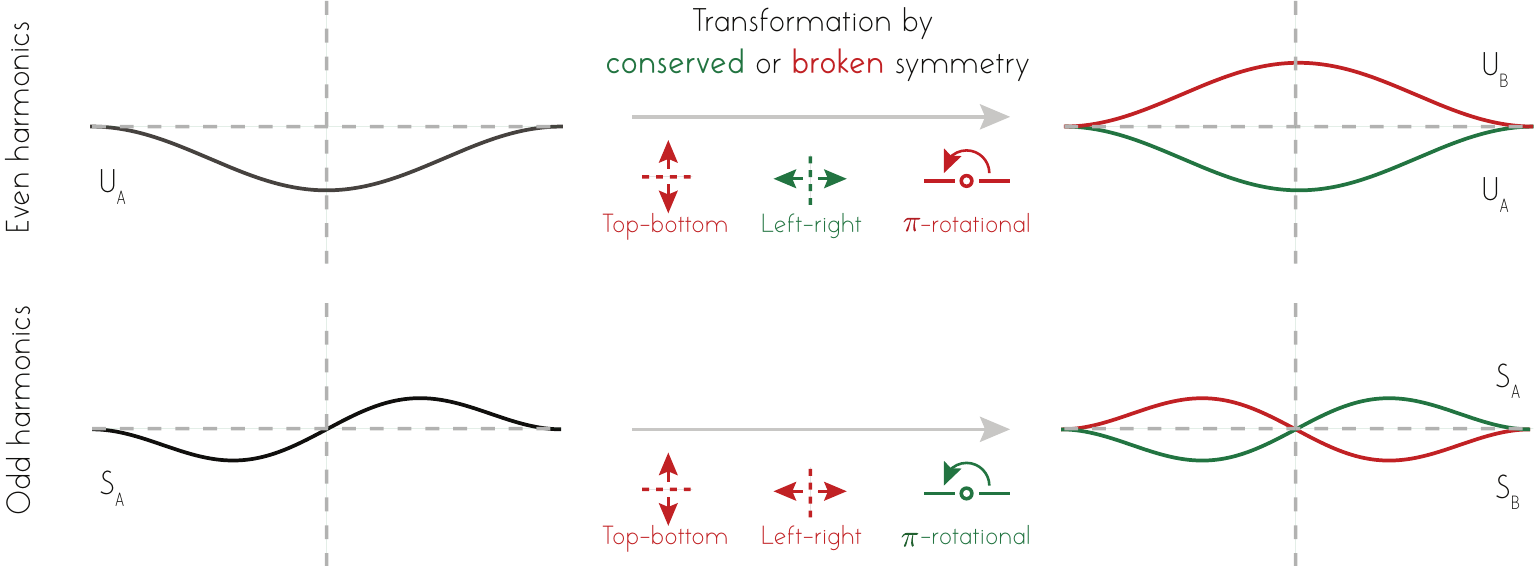}
	\caption{{\textbf{Twin symmetries of the Euler-buckled strip.} The transformation corresponding to a conserved symmetry (green) maps a buckled configuration to itself. The transformation corresponding to a broken symmetry (red) maps a buckled configuration to its twin. We call the $\pi$-rotational symmetry that maps the U-shapes to one another the \textit{U-twin symmetry} and the left-right symmetry that maps the U-shapes to one another the \textit{S-twin symmetry}.}}
	\label{fig:eulerBucklingSymmetries}
\end{figure}

In Euler buckling, the transition from the straight strip equilibrium to any of the buckling harmonics is associated with a spontaneous loss of geometric symmetry of the solution, but not with a loss of invariance of the governing equations of motion. 
%Specifically, for the CC and HH boundary conditions, whenever the strip passes from the straight equilibrium to a buckled state, the top-bottom and either the left-right or the $\pi$-rotational symmetries are lost: the left-right symmetry is lost for all odd harmonics, while the $\pi$-rotational symmetry is lost for all even harmonics. For the CH boundary conditions, the only geometric symmetry available to the straight strip is the top-down symmetry, and this is lost for all buckling harmonics.
The invariance of \eqref{eq:beam_equation_nodim}, \eqref{eq:geometrical_constraint_nodim}, and the CC or HH boundary conditions (Table \ref{tab:dimensional_boundary_conditions}) under the three transformations discussed here guarantees that the image of a solution under any of these transformations is also a solution. 
%Specifically, the image of a solution under a transformation corresponding to a broken geometric symmetry is also a solution. 
%Starting from a buckled equilibrium, if we apply a transformation corresponding to a broken geometric symmetry, we obtain the twin equilibrium; that is, we obtain the other equilibrium in the same pair of equilibria that share the same energy level. This is shown in Fig. \ref{fig:eulerBucklingSymmetries} for the two first harmonics of buckling in the clamped-clamped case. 
In Fig. \ref{fig:eulerBucklingSymmetries}, we consider the equilibrium U\textsubscript{A} (black line) to which we apply all the three transformations. The image of U\textsubscript{A} under the \textit{top-bottom} reflection or \textit{$\pi$-rotation} transformation  is the twin solution U\textsubscript{B} and vice-versa, while the \textit{top-bottom} reflection maps U\textsubscript{A} to itself and U\textsubscript{B} to itself.
Similarly, we show in Fig. \ref{fig:eulerBucklingSymmetries} that the image of S\textsubscript{A} under the \textit{top-bottom} reflection or \textit{left-right} reflection is the twin solution S\textsubscript{B} and vice-versa, while the \textit{$\pi$-rotation} transformation maps S\textsubscript{A} to itself and S\textsubscript{B} to itself.

More generally, the stable U-shapes, as well as the twin shapes of all even harmonics, are related by a $\pi$-rotation about the midpoint of the line connecting the endpoints of the strip. The unstable S-shapes, as well as the twin-shapes of all odd harmonics, are related by a left-right reflection about the line orthogonal to the line connecting the endpoints at its midpoint. We refer to the $\pi$-rotation transformation as the U-twin symmetry and left-right reflection as the S-twin symmetry. The U-twin and S-twin symmetries play an important role in studying shape transitions of the buckled strip under boundary actuation.

%We posit (and we show in the main text) that these symmetries play an important role in understanding shape transitions of the buckled strip under further translational and rotational boundary actuation. 
%Hereafter, we call the transformation that leads to the twin solution that is \textit{unique} to the even and odd harmonics of buckling a \textit{twin transformation}.  Harmonics of buckling in the CC or HH boundary conditions are related by a \textit{twin transformation}: even harmonics are related to each other via the $\pi$-rotational transformation, which we hereafter call the \textit{U-twin symmetry}, and odd harmonics are related to each other via the left-right transformation, which we hereafter denote the \textit{S-twin symmetry}. 
%%Note that in addition to these twin symmetries, the even and odd harmonics are related to each other via the top-down transformation. We do not 

For the CH boundary condition, the \textit{top-bottom symmetry} is the only symmetry at play, and  it is lost for any harmonic of buckling. The two other symmetries in the CC and HH cases do not exist here, as the boundary conditions break invariance of the governing system under the left-right reflection (see Table \ref{tab:dimensional_boundary_conditions}). There is no \textit{twin transformation} with this set of boundary conditions.

%In the main paper, we show that these \textit{twin transformations} play a leading role in the nature of the shape transition the strip undergoes when it is boundary actuated as described in the next sections.

%%%%%%%%%%%%%%%%%%%%%%%%%%%%%%%%%%%%%

\section{Translational boundary actuation of the buckled strip}
\label{sec:translation}

The buckled strip is now driven through shape transition using the transverse translation of one boundary, which  was realized experimentally in Sano \& Wada \cite{sano2018}.
% while rotational boundary actuation was adopted in the experiments of Gomez \textit{et. al} \cite{gomez2017}. In this section, 
Here, we reproduce and expand the results of~\cite{sano2018} numerically using the Cosserat rod theory and semi-analytically using the Euler beam model.
%In~\S~\ref{sec:rotation}, we numerically and analytically reproduce and expand the results of~\cite{gomez2017}. In~\ref{sec:equiv}, we show that the two actuation strategies are equivalent. 

\bigskip
\par
\begin{figure}[!t]
	\centering
	\includegraphics[width=\linewidth]{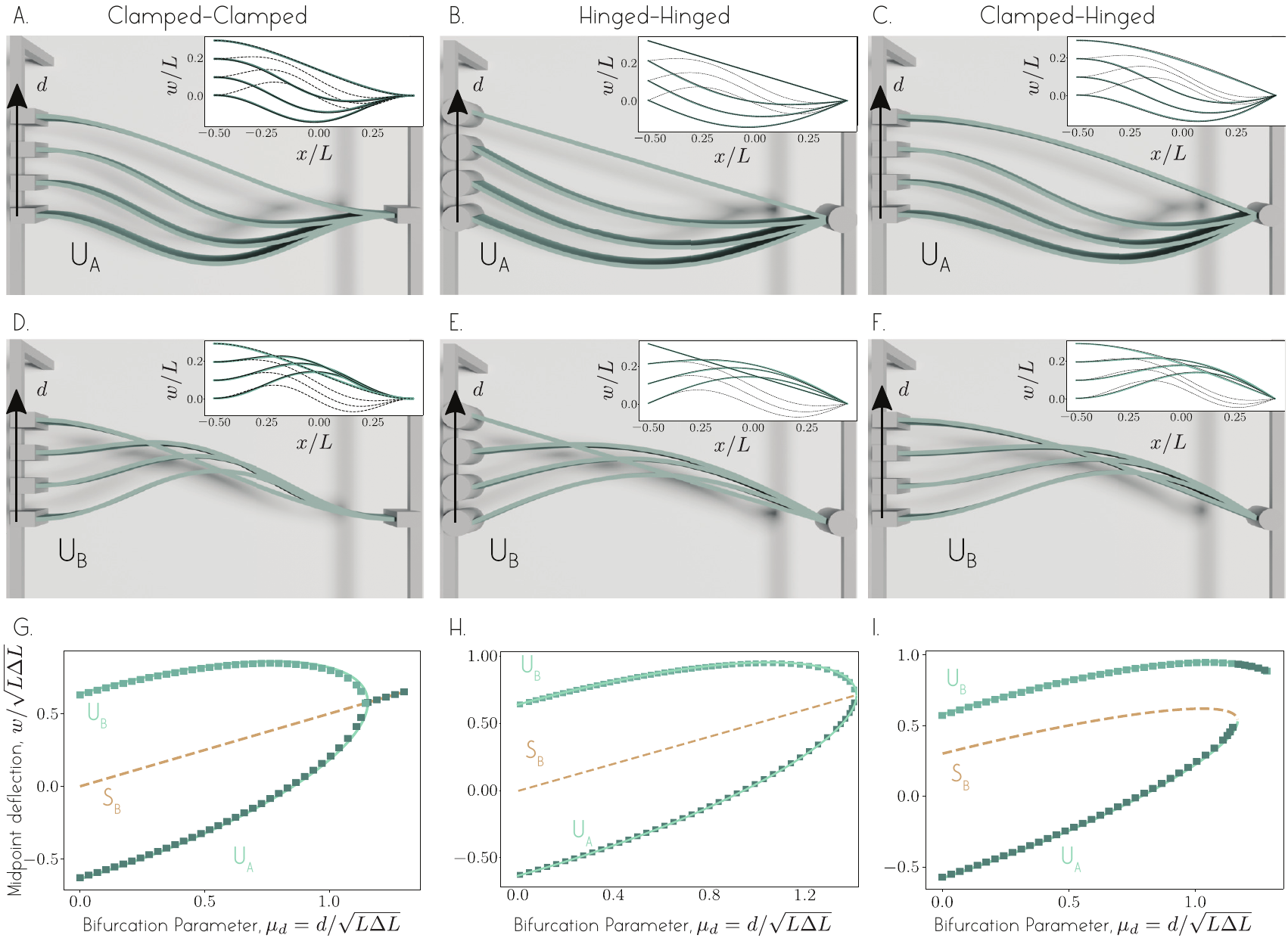}
	\caption{{\textbf{Equilibrium configurations of the Euler-buckled strip under translational actuation.} Starting from the equilibrium shapes U\textsubscript{A,B} of the Euler-buckled, these equilibria morph into different shapes as as a misalignment $d$ is introduced between the two boundaries. \textbf{A-C.} Evolution of the U\textsubscript{A} configuration for different $d$ values for the CC, HH, and CH.  
	\textbf{D-F.} Evolution of the U\textsubscript{B} configuration for different $d$. 
	In the insets in \textbf{A-F}, we compare the equilibrium shapes obtained numerically (green lines show the centerline of the Cosserat rod) to the equilibrium shapes obtained analytically (black lines). \textbf{G-I.}  Evolution of the non-dimensional midpoint deflection of the strip as a function of the non-dimensional misalignment parameter $\mu_d$. The green symbols represent data from numerical Cosserat simulations and the lines data from the Euler beam analysis (full lines for stable equilibrium and dashed lines for unstable ones).}}
	\label{fig:forwardBackwardSW}
\end{figure}

\noindent
\textbf{A. Numerical simulations based on the Cosserat rod theory.}
For each set of boundary conditions, CC, HH, and CH, starting from the stable configuration U\textsubscript{A} of the Euler-buckled strip, we translate the left boundary of the elastic strip in the transverse direction $\mathbf{e}_y$ by an amount $d$ until a maximum value $d_{\textrm{max}}$, allowing the elastic strip to reach mechanical equilibrium at each value $d \in [0,d_{\textrm{max}}]$; see Fig. \ref{fig:forwardBackwardSW} (first row).  
We repeat the same process starting from the other buckled configuration U\textsubscript{B}, see Fig. \ref{fig:forwardBackwardSW} (second row). In all three cases (CC, HH, and CH), bistability is lost beyond a critical value $d^\ast$. The values at which we observe the transition are given in table \ref{tab:muStarNumRot} in terms of the dimensionless bifurcation parameter $\mu = \alpha \sqrt{L/\Delta L}$ discussed below.

\bigskip
\par\noindent
\textbf{B. Equilibria based on Euler beam model.} The boundary conditions due to the transverse misalignment $d$ between the two boundaries
%------
%\begin{equation}
%\left. w\right|_{x=0}=\Delta y, \qquad \left. w\right|_{x=L}=0.
%\label{eq:dim_misalignment}
%\end{equation}
%-------
 can be written in the context of the non-dimensional Euler beam model as (see Table~\ref{tab:dimensional_boundary_conditions}) 
%By virtue of \eqref{eq:nonDimensionnalization}, we rewrite   \eqref{eq:dim_misalignment} in non-dimensional form,
%-----
\begin{equation}
\left. W\right|_{X=-1/2}=\frac{d}{\sqrt{L\Delta L}}\equiv \mu_d, \qquad \left. W\right|_{X=1/2}=0.
\label{eq:adim_misalignment}
\end{equation}
%-----
Here, we introduced the non-dimensional parameter $\mu_d = {d}/{\sqrt{L\Delta L}}$, which balances the vertical position $d$  imposed at the boundary with  the natural vertical position $\sqrt{L\Delta L}$ induced by the end-to-end shortening. 
%This parameter is analogous to the non-dimensional parameter $\mu$ introduced below in \eqref{eq:boundary_smallslope_adim} for rotational actuation, as shown in~\S\ref{sec:equiv}. 

The static equilibria corresponding to the buckled strip driven into translational boundary actuation are obtained from the eigenvalue problem (\ref{eq:linearSystem},\ref{eq:geometrical_constraint_equilibrium}) associated with the non-homogeneous version of~\eqref{eq:linearSystem} accounting for the translational boundary actuation. The three sets of CC, HH, and CH boundary conditions are listed in Table~\ref{tab:dimensional_boundary_conditions}. The CC, HH, and CH boundary conditions give rise to distinct forms of the matrix $\mathbf{M}$, but the translational actuation $\mu_d$ gives rise to the same vector $\mathbf{b}$ on the right hand of~\eqref{eq:linearSystem}. The eigensolutions of the resulting system for each set of boundary conditions are listed in Table~\ref{tab:eigenTranslational}. We note that the form of the solutions in Table~\ref{tab:eigenTranslational} are given for $X\in [0,1]$ as they express in a more compact way than on the interval $X\in [-1/2,1/2]$ adopted in the main paper.

These equilibrium solutions are compared to the equilibria obtained numerically in Fig. \ref{fig:forwardBackwardSW}. In each case, this analysis corroborate our numerical results: there is loss of bistability above a certain threshold $\mu_d=\mu_d^*$. The values of $\mu_d^*$ obtained from the Euler beam analysis are summarized in table \ref{tab:muStarNumRot} along with those obtained from the numerical simulations.

\begin{table}
\caption{\textbf{Bifurcation values} of the control parameters $\mu_d$ and $\mu$ obtained numerically using discrete Cosserat simulations for $\Delta L/L=10^{-2}$ and (semi)-analytically using the Euler beam model. Values without decimal are analytically exact while values with decimals are approximate.}
\begin{tabular}{l|c|c}
\toprule
\multicolumn{3}{c}{$\textbf{TRANSLATIONAL ACTUATION}$}\\
\toprule
 &  Numerical  &
Analytical \\
 & $\mu_d^\ast$  & $\mu_d^\ast$  \\\toprule
$\textbf{Clamped-Hinged:}$ & 1.15 & 1.17\\[1.5ex]
$\textbf{Hinged-Hinged:}$ & 1.40 & $\sqrt{2}$\\[1.5ex]
$\textbf{Clamped-Clamped:}$ & 1.14 & $\sqrt{4/3}$\\[1.5ex]
\hline
\end{tabular}
\hspace{0.5in}
\begin{tabular}{l|c|c}
\toprule
%\multicolumn{3}{c}{}\\[1.5ex]
\multicolumn{3}{c}{$\textbf{ROTATIONAL ACTUATION}$}
\\
\toprule
 & Numerical  & 
Analytical \\
 & $\mu^\ast$  & $\mu^\ast$  \\\toprule
$\textbf{Asymmetric:}$  & 1.763 & 1.782 \\[1.5ex]
$\textbf{Symmetric:}$  & 1.973 & 2 \\[1.5ex]
$\textbf{Antisymmetric:}$  & 1.967 & 2 \\[1.5ex]
\hline
\end{tabular}
\label{tab:muStarNumRot}
\end{table}

%%%%%%%%%%%%%%%%%%%%%%%%%%%%%%%%%%%%%%%%%%%%%%%%%%%%%%%%%%%%%%%%%%
\section{Rotational boundary actuation of the buckled elastic strip}
\label{sec:rotation}

\begin{figure}
	\centering
	\includegraphics[width=\linewidth]{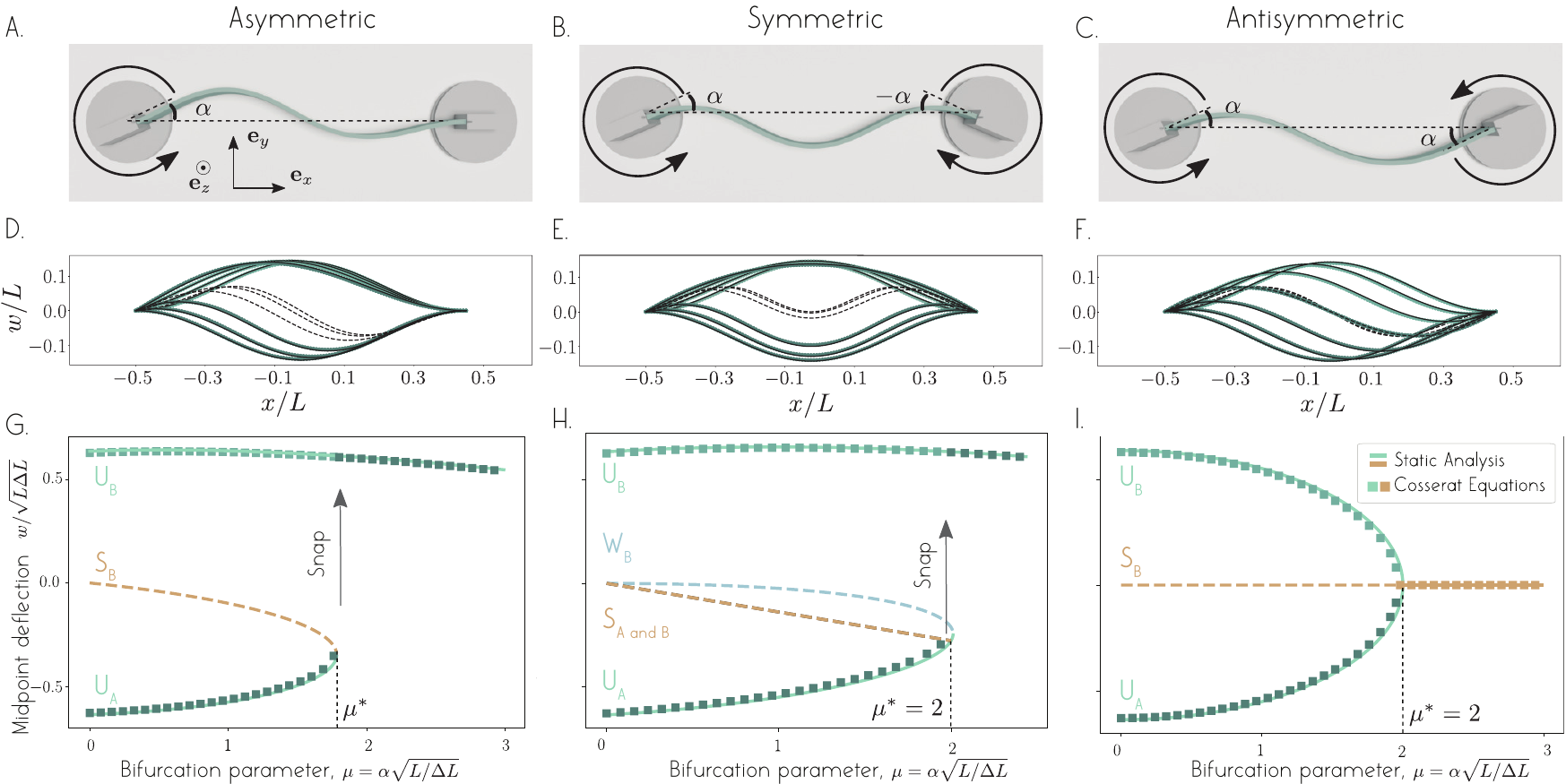}
	\caption{\textbf{Equilibrium configurations of the Euler-buckled strip under rotational actuation.}  One or both ends of the CC strip are (quasi-statically) rotated  by an angle $\alpha$, leading to loss of bistability as $\alpha$ increases.  \textbf{A,B.}  Asymmetric actuation (one end is rotated) and symmetric actuation (both ends are rotated by the same amount in opposite direction) lead to violent snap-through. \textbf{C.} Antisymmetric actuation (both ends are rotated by the same amount in the same direction) leads to a smooth transition. \textbf{D-F.} Evolution of the shapes U\textsubscript{A}, U\textsubscript{B} and S\textsubscript{B} for different values of $\mu$. The green lines represent the centerline of the Cosserat rod. The black lines represent the equilibrium shapes obtained from the static analysis of the Euler beam model. \textbf{G-I.} Midpoint deflection of the strip as a function of the bifurcation parameter $\mu = \alpha \sqrt{L/\Delta L}$. Green squares represent data obtained from numerical simulations based on the Cosserat rod theory. Solid and dashed lines represent, respectively, stable and unstable branches obtained from the static analysis of the Euler beam model.}
	\label{fig:forwardBackwardGMV}
\end{figure}

We numerically and analytically reproduce and expand the results of~\cite{gomez2017} for a clamped-clamped buckled strip driven into rotational boundary actuation. Starting from the clamped-clamped buckled strip, we actuate the strip by rotating one or both ends. 
Specifically, we consider three types of rotational boundary actuation: asymmetric where only one end is rotated by an angle $\alpha$,  symmetric where both ends are rotated by the same angle $\alpha$ in two opposite directions, and  antisymmetric where the two ends are rotated by the same angle $\alpha$ in the same direction (see Fig.~\ref{fig:forwardBackwardGMV}). 

\bigskip
\par\noindent
\textbf{A. Numerical simulations based on the Cosserat rod theory.}
For each type of boundary actuation, we start from the initial configuration U\textsubscript{A} and we increase $\alpha$ by small increments $\Delta\alpha$ until a maximum $\alpha_{\textrm{max}}$ is reached, allowing the elastic strip to reach an equilibrium configuration at each value $\alpha \in [0,\alpha_{\textrm{max}}]$. We repeat the same process starting from the initial configuration U\textsubscript{B}. Representative equilibrium configurations obtained for select $\alpha$ values are shown in Fig.~\ref{fig:forwardBackwardGMV}D,~\ref{fig:forwardBackwardGMV}E,~\ref{fig:forwardBackwardGMV}F.
In all three types of boundary actuation, bistability is lost after at a critical value $\alpha^\ast$.  The values at which we observe the transition are given in table \ref{tab:muStarNumRot} in terms of the dimensionless bifurcation parameter $\mu = \alpha \sqrt{L/\Delta L}$ discussed below.

\bigskip
\par\noindent
\textbf{B. Equilibria based on Euler beam model.} 
The boundary conditions for asymmetric, symmetric, and antisymmetric actuation are given in Table~\ref{tab:dimensional_boundary_conditions} in the limit of small angle $\alpha \ll 1$. Specifically, the boundary conditions in terms of the dimensionless parameter $\mu = \alpha\sqrt{L/\Delta L}$ of the strip at the rotated end(s) are
%-------
\begin{equation}
\textrm{asymmetric:} \left.\frac{\partial W}{\partial X}\right|_{X=-1/2}=
\mu, 
\qquad 
\textrm{symmetric:} \left.\frac{\partial W}{\partial X}\right|_{X=-1/2,1/2}= \pm\mu,
\qquad
\textrm{antisymmetric:} \left.\frac{\partial W}{\partial X}\right|_{X=-1/2,1/2}= \mu.
%\left.\frac{\partial W}{\partial X}\right|_{X=0\ \text{or}\ X=1}=\sqrt{\frac{L}{\Delta L}}\alpha\equiv \mu
\label{eq:boundary_smallslope_adim}
\end{equation}
%-------
The non-dimensional bifurcation parameter $\mu = \alpha \sqrt{{L}/{\Delta L}}$, first introduced by Gomez et al. \cite{gomez2017,gomez2018}, 
balances the slope $\alpha$ imposed at the boundary with the natural slope $\sqrt{\Delta L/L}$ adopted by the buckled strip.
 If $\alpha$ remains small compared to $\sqrt{\Delta L/L}$ ($\mu\ll1$), the angle imposed at the boundary has a small influence on the overall shape of the strip compared to the influence of the longitudinal geometrical constraint. If on the other hand $\mu\sim O(1)$, the shape of the strip will be influenced both by the end-to-end shortening and by the angle imposed at the boundaries.

The static equilibria corresponding to the buckled strip driven into rotational boundary actuation are obtained from the eigenvalue problem associated with the non-homogeneous version of~\eqref{eq:linearSystem} accounting for the boundary conditions in~\eqref{eq:boundary_smallslope_adim}. 
Each set of boundary conditions (asymmetric, symmetric, and antisymmetric) in Table~\ref{tab:dimensional_boundary_conditions} gives rise to a vector $\mathbf{b}$ on the right hand of~\eqref{eq:linearSystem}.  (Semi)-analytic solutions to this eigenvalue problem are summarized in Table \ref{tab:eigenRotational}. 

These equilibrium solutions are compared to the equilibria obtained numerically in Fig. \ref{fig:forwardBackwardGMV}. In each case, this analysis corroborate our numerical results: there is loss of bistability above a certain threshold $\mu=\mu^*$. The values of $\mu^*$ obtained from the Euler beam analysis are summarized in table \ref{tab:muStarNumRot} along with those obtained from the numerical simulations.

%%%%%%%%%%%%%%%%%%%%%%%%%%%%%%%%%%%%%%%%%%%%%%%%%%%%%%%%%%%%
\section{Geometric symmetry breaking and its effect on the eigenvalue problem}
\label{sec:eigenstructure}

The families of equilibrium solutions obtained under translational and rotational actuation of the strip's boundaries can be compared to the families of equilibria of the Euler-buckled strip. Of particular interest is the role of the translational and rotational boundary actuation in breaking the twin symmetries identified in the Euler-buckled strip (see \S\ref{sec:buckling}). Four categories emerge from this comparison.

Consider first the translational actuation and antisymmetric rotational actuation of the clamped-clamped strip. In these cases, the non-homogeneous system admits two families of solutions: one is inherited from the even harmonics found in the homogeneous system, the other is obtained by inverting \eqref{eq:linearSystem}. This means that, for these two sets of boundary conditions, the right hand side $\mathbf{b}$ lies in the subspace described by the family of even harmonics of the homogeneous problem. This is a mathematical illustration of the fact that the $\pi$\textit{-rotational twin symmetry} associated with the even buckling harmonics is satisfied by this actuation. There is no family of solution inherited from the odd harmonics of the homogeneous system because the right-hand side $\mathbf{b}$ in \eqref{eq:linearSystem} breaks the \textit{left-right twin symmetry} associated with this family of solutions.

Consider now the translational actuation of the hinged-hinged strip. The non-homogeneous system admits solutions for all values of $\Lambda$ for which $\det(\mathbf{M})=0$. This means that $\mathbf{b}$ lies in the subspace described by the even and odd harmonics of buckling of the homogeneous system. This is a mathematical illustration of the fact that this actuation satisfies the twin symmetry of all buckled equilibria associated with the homogeneous problem. Indeed as the strip is hinged on both sides, the two boundaries remain always aligned with the line joining the two boundaries and therefore satisfy a reflection symmetry about the line connecting the strip's endpoints.

Next consider the case of symmetric rotational actuation in which the non-homogeneous system admits two families of solutions:  one is obtained by inverting \eqref{eq:linearSystem} and the other is inherited from the odd harmonics of buckling of the homogeneous system. This means that the right hand side $\mathbf{b}$ lies in the subspace described by odd harmonics of the the homogeneous system. This is a mathematical illustration of the fact that this set of boundary conditions does not break the \textit{left-right twin symmetry} associated with the odd harmonics of buckling. The family corresponding to the even harmonics of the homogeneous system is not inherited because the right-hand side $\mathbf{b}$ breaks the $\pi$\textit{-rotational twin symmetry} associated with this family of solutions.
 
Lastly, consider the case of translational actuation of the clamped-hinged strip and the asymmetric rotational actuation of the clamped-clamped strip. Both these configurations admit only one family of solution obtained by inverting \eqref{eq:linearSystem}. There is no family of solution inherited from the family of solutions of the homogeneous system. This means that the solution corresponding to values of $\Lambda$ for which $\det(\mathbf{M})$ vanishes are no longer solutions to the non-homogeneous system in \eqref{eq:linearSystem}. That is, the right hand side $\mathbf{b}$ does not lie in the subspace described by the solutions of the homogeneous system. This is a mathematical illustration of the fact that this set of boundary condition breaks all the twin symmetries of the Euler buckling problem.

%%%%%%%%%%%%%%%%%%%%%%%%%%%%%%%%%%%%%%%%%%%%%
\section{Force measurements in the case of translational boundary actuation}

In this section, we carry a discussion about the force measured by Sano \& Wada~\cite{sano2018} for the translationally actuated strip, and we reinterpret their observations based on the symmetry breaking mechanism introduced in the main text.  Sano \& Wada~\cite{sano2018} measured the transverse force (in the $y$ direction) applied by the beam on the left clamped boundary (at $X=0$). They observed that in the clamped-hinged case, the force plotted as a function of the bifurcation parameter $\mu_d$ exhibits a hysteresis depending on whether the beam is in the U\textsubscript{A} or U\textsubscript{B} configuration. Surprisingly, this hysteresis is not observed in the clamped-clamped and hinged-hinged cases. This hysteresis is actually a consequence of the symmetry breaking mechanism unravelled in the main paper.

\begin{figure}[!t]
	\centering
	\includegraphics[width =\textwidth]{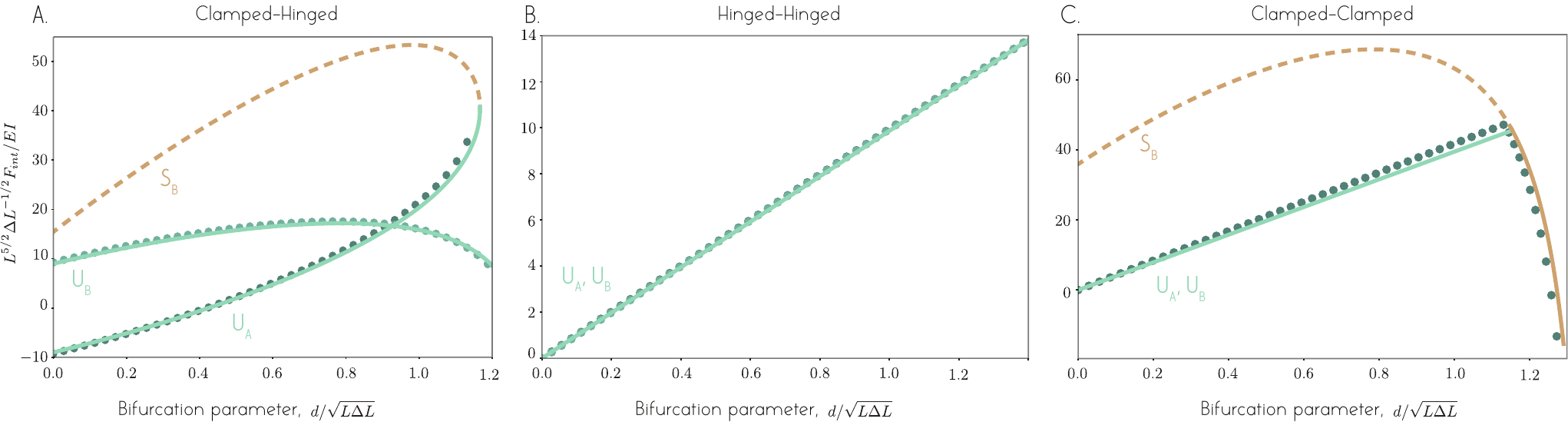}
	\caption{{\textbf{Transverse force applied by the strip on the left boundary.} Evolution of the non-dimensional transverse force applied  on the left clamped boundary of the strip under translational actuation (see Fig.~\ref{fig:forwardBackwardSW}) in term of the  non-dimensional bifurcation parameter $\mu_d$. The force is obtained analytically from the Euler beam model (full line) and numerically from the discrete Cosserat equations (symbols).}}
	\label{fig:internalForce}
\end{figure}

Here, we exploit both our numerical simulations and the Euler beam model to compute this force. 

In our simulations, the force is obtained by "measuring" the force applied by the beam on the boundary, as done experimentally in \cite{sano2018}. The obtained values are reported as dots on Fig. \ref{fig:internalForce}.

In the Euler-beam model, the non-dimensional transverse force $F_{\textrm{int}}(X)$ applied at abscissa $X$ along the strip is given  by
%---------
\begin{equation}
	F_{\textrm{int}}(X)=-\frac{\partial^3 W}{\partial X^3}-\Lambda^2\frac{\partial W}{\partial X}.
	\label{eq:internal_force}
\end{equation}
%----------
where the first term in the right hand side corresponds to bending forces and the second term to compression forces. We substitute the analytical shape of the strip at equilibrium (see Table~\ref{tab:eigenTranslational}) into~\eqref{eq:internal_force} to calculate the transverse force applied at the left end of the strip for each of the three cases: CH, HH, and CC.

%\subsection{Clamped-Hinged case :}
\bigskip
\par\noindent
\textbf{A. Clamped-Hinged case.}
We compute the transverse force for the case $d=0$ and $d \neq 0$ separately.
 When the misalignment is zero ($d=0$),  the transverse force  \eqref{eq:internal_force} is computed using the expression of $W(X)$ in Table~\ref{tab:eigenEulerBuckling} for U\textsubscript{A} and U\textsubscript{B}
 %---------
\begin{equation}
%\left\{
\begin{split}		
F_{\textrm{int}}^{\textrm{U}_\textrm{A}}=-F_{\textrm{int}}^{\textrm{U}_\textrm{B}} = \displaystyle{\frac{2 \sqrt{2} \Lambda^3}{\sqrt{\Lambda \left(2 \Lambda^3-\Lambda^2 \sin (2 \Lambda)-8 \sin (\Lambda)+\sin (2 \Lambda)+8 \Lambda \cos (\Lambda)-2 \Lambda \cos (2 \Lambda)\right)}}},
%\\[3mm]
%F_{\textrm{int}}^{\textrm{U}_B}&=&\displaystyle-\frac{2 \sqrt{2} \Lambda^3}{\sqrt{\Lambda \left(2 \Lambda^3-\Lambda^2 \sin (2 \Lambda)-8 \sin (\Lambda)+\sin (2 \Lambda)+8 \Lambda \cos (\Lambda)-2 \Lambda \cos (2 \Lambda)\right)}}.\\[3mm]
\end{split}
%\right.
\end{equation}
%-----------
Here, $F_{\textrm{int}}^{\textrm{U}_\textrm{A}}$  is the transverse force applied by the strip on the left boundary when the strip is in the U\textsubscript{A}  configuration. Similarly, $F_{\textrm{int}}^{\textrm{U}_\textrm{B}}$ is the transverse force when the strip is in the U\textsubscript{B} configuration.
The force is non-zero for $d=0$ and depends on which side the strip buckles, U\textsubscript{A} or U\textsubscript{B}. This is a consequence of the asymmetry between the clamped and hinged boundary conditions. The moment applied on the strip by the left clamped boundary induces a bending moment in the strip that cannot be balanced by other means than by applying a transverse force at the right hinged end (because there is no moment at a hinged end) that in turn must be balanced by a transverse force at the left end in order to guarantee mechanical equilibrium of the system. 

For $d\neq 0$, the transverse force \eqref{eq:internal_force} is computed from the equilibrium expressions for U\textsubscript{A} and U\textsubscript{B} given in Table~\ref{tab:eigenTranslational}. We obtain the same expression for both U\textsubscript{A} and U\textsubscript{B}:
%---------
\begin{equation}
    F_{\textrm{int}}^{\textrm{U}_A}=F_{\textrm{int}}^{\textrm{U}_B}=\frac{\Lambda^3 \cos (\Lambda)}{\Lambda \cos (\Lambda)-\sin (\Lambda)}\frac{d}{\sqrt{L\Delta L}}.
    \label{eq:internalForceCH}
\end{equation}
%-------------
This means that the transverse force $F_{\textrm{int}}^{\textrm{U}_A}$ and $F_{\textrm{int}}^{\textrm{U}_B}$ would be the same only if the eigenvalue $\Lambda$ associated with U\textsubscript{A} is the same as the one associated with U\textsubscript{B}. This is not the case because of the symmetry breaking mechanism identified in the main paper. As soon as $d\neq0$, U\textsubscript{A} and U\textsubscript{B} are no longer energetically equivalent (see main text) and their eigenvalue $\Lambda$ is no longer equal. This explains the strong hysteresis observed in \cite{sano2018}.

%This force would be the same for the two buckled modes only if the tension $\Lambda$ is of the same magnitude for the two solutions, which is not the case.
%This asymmetry in the transverse force can be interpreted in terms of the symmetry breaking mechanism discussed in the main text. First, at $\mu=0$, the left-right symmetry is broken by introducing clamped-hinged boundary conditions, and the transverse force is forced to be non-zero to satisfy equilibrium and balance the bending moment induced by the clamped end. As the average curvature along the beam is reversed between U\textsubscript{A} and U\textsubscript{B}, the direction of the bending moment and consequently the direction of the transverse force is also reversed making this force shape-dependent. 
%When a misalignment $d$ is applied, the transverse force would kept the same magnitude (and opposite signs) at U\textsubscript{A}, U\textsubscript{B} only if the twin symmetry associated with U\textsubscript{A}, U\textsubscript{B} were conserved, which is not the case \ek{I'm not sure I follow your logic here}. As soon as $d \neq 0$, the shape is no longer the same for the two configurations, implying different bending moments and consequently different shearing forces which explains the strong hysteresis observed by Sano \& Wada~\cite{sano2018}. 
In Fig.~\ref{fig:internalForce}A, we plot the analytically obtained transverse force for the U\textsubscript{A}, U\textsubscript{B} and S\textsubscript{B} configurations (solid lines) on top of the numerical values obtained from the Cosserat model (green symbols). The analytical branches (solid lines) compare well with the numerical values (green symbols). Our data confirm the strong hysteresis observed experimentally by Sano \& Wada~\cite{sano2018}.

\bigskip
\par\noindent
\textbf{B. Hinged-Hinged Case.}
 We calculate the transverse force from \eqref{eq:internal_force} using the equilibrium shapes U\textsubscript{A} and U\textsubscript{B} for the hinged-hinged setup (see  Table~\ref{tab:eigenTranslational}),
 %----------
\begin{equation}
F_{\textrm{int}}^{\textrm{U}_\textrm{A}}=F_{\textrm{int}}^{\textrm{U}_\textrm{B}}=\Lambda^2\frac{d}{\sqrt{L\Delta L}}.
\label{eq:internal_force_HH1}
\end{equation}
%-------------
In contrast to the clamped-hinged case, the expression for the transverse force is independent of shape. When $d=0$, the transverse force associated with bending (first term in the rhs of \eqref{eq:internal_force}) is balanced by the transverse force due to tension in the beam (second term in the rhs of \eqref{eq:internal_force}) such that the resultant is zero. When $d \neq 0$ the part of the transverse force that is not balanced by the bending force is the part due to the compression force $\Lambda^2$ acting in the direction of the line connecting the two hinged boundaries. The projection of this force along $\mathbf{e}_y$ yields \eqref{eq:internal_force_HH1}. 
Thus, there is no hysteresis because the transverse force measured along $\mathbf{e}_y$ in $X=0$ is not shape dependent but only due to the compression force $\Lambda^2$ that remains aligned with the line connecting the two boundaries and is thus no longer perpendicular to $\mathbf{e}_y$.

In Fig. \ref{fig:internalForce}B, we plot the force in \eqref{eq:internal_force_HH1} as a function of the bifurcation parameter $\mu_d$ (solid lines) and we compare it to the value of the force at the left boundary obtained numerically (green symbols) for both  U\textsubscript{A} and U\textsubscript{B}. The transverse force obtained numerically compares well with the analytical branches and they are both exactly the same for U\textsubscript{A} and U\textsubscript{B}, confirming the results obtained by Sano \& Wada~\cite{sano2018}. 

\bigskip
\par\noindent
\textbf{C. Clamped-Clamped Case.}
As $d$ increases, the strip transitions from the U\textsubscript{A, B} configuration to the S\textsubscript{B} configuration. We compute the transverse force associated with U\textsubscript{A, B} and S\textsubscript{B} separately.
%\paragraph{U\textsubscript{A,B} configurations:}
Consider the expression $W(X)$ of the  U\textsubscript{A} and U\textsubscript{B} configurations given in Table~\ref{tab:eigenTranslational}, and substitute this expression in \eqref{eq:internal_force}. We obtain the expression for the transverse force,
%------
\begin{equation}
    F_{\textrm{int}}^{\textrm{U}_\textrm{A}}=F_{\textrm{int}}^{\textrm{U}_\textrm{B}}=\Lambda^2\frac{d}{\sqrt{L\Delta L}}.
    \label{eq:internal_force_CC1}
\end{equation}
%-------
Here, we have analytical evidence of the symmetric property of the transverse force, observed in \cite{sano2018} in a numerical experiment. Similarly to the Hinged-Hinged case, it is surprising that, although the shape is reversed, the force applied by the beam on the boundary is the same in the U\textsubscript{A} and U\textsubscript{B} configuration.  When the two clamped boundaries are aligned ($d=0$), the transverse force is zero. The forces associated with the bending moments in the beam (first term in \eqref{eq:internal_force}) exactly equals the forces induced by the compression force (second term in \eqref{eq:internal_force}). When a misalignment is applied ($d\neq0$) between the two boundaries, the compression force $\Lambda^2$ acting in the direction of the line connecting the two clamped boundaries is no longer perpendicular to $\mathbf{e}_y$ and its projection on $\mathbf{e}_y$, yields \eqref{eq:internal_force_CC1}.  Therefore, the transverse force in the beam is independent of the shape of the beam and only depends on the misalignment between the two boundaries. This explains why, for a given $\mu_d$ value, the forces are exactly the same in the U\textsubscript{A} and U\textsubscript{B} configurations.

For the S\textsubscript{B} configuration, the transverse force takes the form
%---------
\begin{equation}
      F_{\textrm{int}}=\frac{\Lambda^3\cos(\Lambda/2)}{\Lambda\cos(\Lambda/2)-2\sin(\Lambda/2)}\frac{d}{\sqrt{L\Delta L}}.
      \label{eq:internal_force_CC2}
\end{equation}
%-----------
In Fig. \ref{fig:internalForce}C, we show the two branches of solution for the U\textsubscript{A, B} and S\textsubscript{B} configurations and we compare these results to values of the force applied by the beam on the left boundary obtained numerically. The analytical branches compare well with the numerical values. These forces are exactly the same for U\textsubscript{A} and U\textsubscript{B},  confirming the result obtained by Sano \& Wada~\cite{sano2018}.

%%%%%%%%%%%%%%%%%%%%%%%%%%%%%%%%%%%%%%%%%%%%%%%%%%%%
\section{Equivalence between translational and rotational boundary actuation}
\label{sec:equiv}

We show that translational actuation of the Euler-buckled strip, implemented experimentally in~\cite{sano2018} and analyzed in detail in section~\ref{sec:translation}, is equivalent to the rotational actuation of the Euler-buckled strip, studied experimentally and analytically in~\cite{gomez2017} and analyzed in detail in section~\ref{sec:rotation}. 

\begin{figure}[!t]
	\centering
	\includegraphics[width =\textwidth]{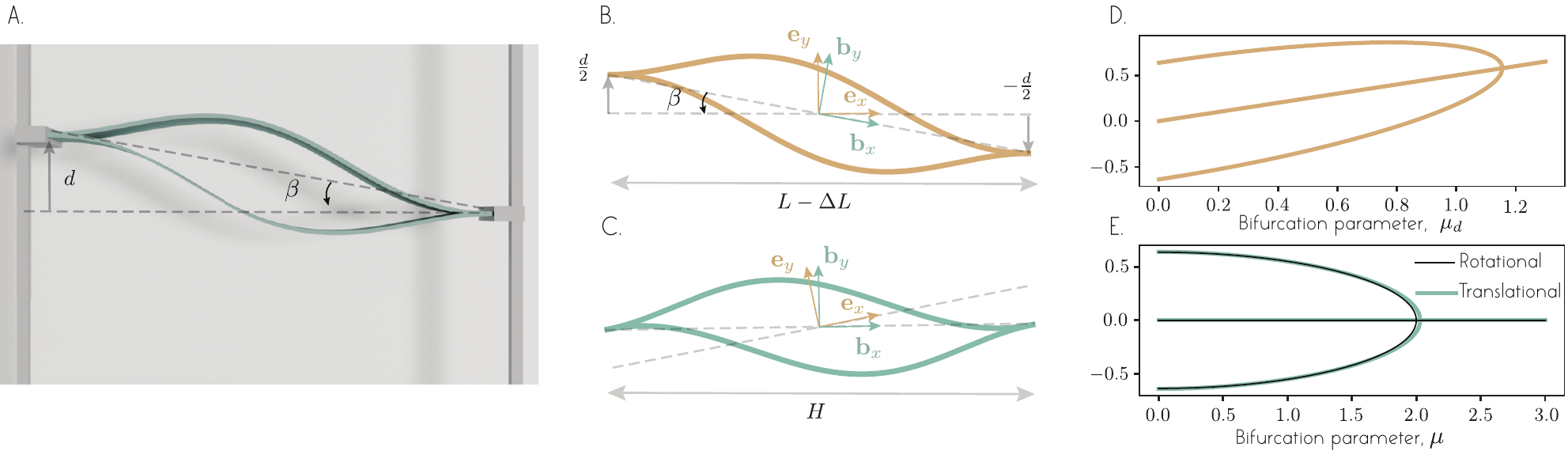}
	\caption{{\textbf{Lagrangian frame of reference} If we observe the system from a frame attached to the line that connects the two ends of the strip, the translational actuation becomes a rotational actuation. \textbf{A.} 3D rendering of the two equilibria U\textsubscript{A} and U\textsubscript{B} obtained numerically with the t-CC actuation and observed from the Eulerian frame $(\mathbf{e}_x, \mathbf{e}_y, \mathbf{e}_z)$. \textbf{B.} Analytical equilibria obtained for the t-CC actuation, depicted in the $(\mathbf{e}_x, \mathbf{e}_y, \mathbf{e}_z)$ Eulerian frame. \textbf{C.} Analytical equilibria obtained for the t-CC actuation, depicted in the $(\mathbf{b}_x, \mathbf{b}_y, \mathbf{b}_z)$ Lagrangian frame. \textbf{D.} Bifurcation diagram obtained analytically for the t-CC case. \textbf{E.} Comparison of the bifurcation diagram obtained for the t-CC case (green lines) and expressed in the Lagrangian frame $(\mathbf{b}_x, \mathbf{b}_y, \mathbf{b}_z)$ with the bifurcation diagram obtained for the antisymmetric case (black lines).}}
	\label{fig:referentialOfTheBeam}
\end{figure}

We introduce a Lagrangian frame of reference $(\mathbf{b}_x, \mathbf{b}_y, \mathbf{b}_z)$  whose $x$ axis remains aligned with the line joining the two ends of the strip (Fig. \ref{fig:referentialOfTheBeam}B). For the translational actuation, a mapping from the Eulerian frame $(\mathbf{e}_x, \mathbf{e}_y, \mathbf{e}_z)$ to this new frame of reference corresponds to a rigid body rotation by an angle $\beta=-\arctan(\Delta y / (L-\Delta L))$ about the $\mathbf{e}_z$ axis. The dimensional coordinates of any point $\mathbf{r}$ in space can thus be transformed from the Eulerian frame to the Lagrangian frame using the rigid-body rotation defined by
%-------
\begin{equation}
\mathbf{R}=\begin{pmatrix}
\cos\beta & \sin\beta & 0\\
-\sin\beta & \cos\beta & 0\\
0&0&1
\end{pmatrix}
\label{eq:matrixChangeOfFrame}
\end{equation}
%--------

Consider the case of translational actuation of the clamped-clamped strip. In the Eulerian frame $(\mathbf{e}_x, \mathbf{e}_y, \mathbf{e}_z)$, the boundary conditions in dimensional form are given by (see Fig. \ref{fig:referentialOfTheBeam}B)
%-------
\begin{equation}
\left.w\right|_{x=-\frac{1}{2}(L-\Delta L)}=\frac{\Delta y}{2},
\qquad
\left.\frac{\partial w}{\partial x}\right|_{x=-\frac{1}{2}(L-\Delta L)}=0,
\qquad 
\left.w\right|_{x=\frac{1}{2}(L-\Delta L)}=-\frac{\Delta y}{2},
\qquad
\left.\frac{\partial w}{\partial x}\right|_{x=\frac{1}{2}(L-\Delta L)}=0.
\label{eq:bcsEulerian}
\end{equation}
%--------
The boundary conditions in the Lagrangian frame $(\mathbf{b}_x, \mathbf{b}_y, \mathbf{b}_z)$, see Fig. \ref{fig:referentialOfTheBeam}C, are obtained by applying the rigid-body rotation in~\eqref{eq:matrixChangeOfFrame} to~\eqref{eq:bcsEulerian},
%---------
\begin{equation}
\left.w\right|_{x=-H/2}=0,
\qquad
\left.\frac{\partial w}{\partial x}\right|_{x=-H/2}=\tan\beta,
\qquad 
\left.w\right|_{x=H/2}=0,
\qquad
\left.\frac{\partial w}{\partial x}\right|_{x=H/2}=\tan\beta.
\label{eq:bcsLagrangian}
\end{equation}
%----------
Here, we introduced $H=\sqrt{\Delta y^2+(L-\Delta L)^2}$. The new set of boundary conditions \eqref{eq:bcsLagrangian} shows that, once observed in the proper frame of reference, the translational actuation is equivalent to a rotation of the boundaries by an angle $\beta$ while pulling the boundaries away from each others. 

To find the corresponding non-dimensional angle applied at the boundaries, we need to identify the natural horizontal and vertical length scales in the new frame of reference. According to \eqref{eq:bcsLagrangian}, we take $H$ to be the horizontal length scale. Following \cite{gomez2017}, the natural vertical length scale is obtained by expressing the horizontal confinement.
%-------
\begin{equation}
\int_{-L/2}^{L/2}\cos\left(\theta(s)\right)ds=H
\label{eq:horizontalConfinement}
\end{equation}
%--------
where $\theta(s)$ is the local angle made by the beam with $\mathbf{b}_x$ at the curvilinear coordinate $s$ along the beam. Writing \eqref{eq:horizontalConfinement} in the limit $\theta\ll 1$ we find the natural vertical length scale $w\sim\sqrt{H(L-H)}$.
Thus, once in the Lagrangian frame of reference, we  non-dimensionalized the quantities $w$ and $x$ using
%------
\begin{equation}
X=\frac{x}{H}\qquad W=\frac{w}{\sqrt{H(L-H)}},
\label{eq:adimQuantitiesNewRef}
\end{equation}
%--------
and obtain the corresponding non-dimensional angle imposed at the boundaries
%----------
 \begin{equation}
\mu=\frac{\Delta y}{L-\Delta L}\sqrt{\frac{H}{L-H}}.
\label{eq:adimAngleNewRef}
\end{equation}
%-----------

The transformation \eqref{eq:matrixChangeOfFrame} and the non-dimensionalization \eqref{eq:adimQuantitiesNewRef} allow us to map the bifurcation diagram associated with the translational actuation (Fig. \ref{fig:referentialOfTheBeam}D) into a new bifurcation diagram corresponding to a rotational actuation in the new frame of reference (Fig. \ref{fig:referentialOfTheBeam}E green lines). The comparison of this new bifurcation diagram with the bifurcation diagram obtained in the case of antisymmetric rotational actuation (Fig. \ref{fig:referentialOfTheBeam}E black lines) shows that translational actuation of the clamped-clamped strip is equivalent to antisymmetric rotational actuation of the strip. The only difference being that, in the case of antisymmetric rotational actuation, $\mu$ is increased by varying the angle imposed at the boundaries whereas, according to \eqref{eq:adimAngleNewRef}, in the translational actuation, $\mu$ is increased by increasing the angle $\beta$ imposed at the boundaries while decreasing the end-to-end confinement.

We now apply the change of frame of reference in~\eqref{eq:matrixChangeOfFrame} and the non-dimensionalization in~\eqref{eq:adimQuantitiesNewRef} to the case of translational actuation of the hinged-hinged strip. With hinged boundaries, there is no torque applied by the boundaries at the endpoints and the boundary conditions in the Lagrangian frame become
%--------
\begin{equation}
\left.w\right|_{x=-H/2}=0,
\qquad
\left.\frac{\partial^2 w}{\partial x^2}\right|_{x=-H/2}=0,
\qquad 
\left.w\right|_{x=H/2}=0,
\qquad
\left.\frac{\partial^2 w}{\partial x^2}\right|_{x=H/2}=0.
\label{eq:bcsLagrangiantHH}
\end{equation}
%--------
Thus, the problem is homogeneous in this reference frame. This corresponds to the Euler buckling problem studied in \ref{sec:buckling}, except that here, the two boundaries are being pulled apart (instead of being pushed towards each others) until the strip gets back to its straight configuration. This actuation, therefore satisfies all the twin symmetries of the Euler buckling problem. This explains that the non-homogeneous system solved in \ref{sec:buckling} for this case admits solutions for all the eigenvalues obtained in the homogeneous case, as discussed in section \ref{sec:eigenstructure}.

Lastly, we apply the change of frame of reference in~\eqref{eq:matrixChangeOfFrame} and the non-dimensionalization in~\eqref{eq:adimQuantitiesNewRef} to the case of translational actuation of the clamped-hinged strip. We obtain the following set of boundary conditions
%---------
\begin{equation}
\left.w\right|_{x=-H/2}=0,
\qquad
\left.\frac{\partial w}{\partial x}\right|_{x=-H/2}=\tan\beta,
\qquad 
\left.w\right|_{x=H/2}=0,
\qquad\left.\frac{\partial^2 w}{\partial x^2}\right|_{x=H/2}=0.
\label{eq:bcsLagrangiantCH}
\end{equation}
%---------
This case corresponds to a clamped-hinged strip actuated by rotating the left boundary by an angle $\beta$ while pulling the two boundaries away from each others. Therefore, this actuation breaks the twin symmetry of both the U and S shape and exhibit a saddle-node bifurcation as the asymmetric case studied in \cite{gomez2017} and in the main paper.

%%%%%%%%%%%%%%%%%%%%%%%%%%%%%%%%%%%%%%%%%%%%%%%%%%%%%%%

%%%%%%%%%%%%%%%%%%%%%%%%%%%%%%%%%%

%%%%%%%%%%%%%%%%%%%%%%%%%%%%%%%%%%%

\section{Methods for plotting bifurcation diagrams}\label{sec:bifurcationDiagrams}

In Figs.~1, 2 and 4 of the main text, as well as in Figs.~\ref{fig:forwardBackwardSW}, ~\ref{fig:forwardBackwardGMV} and ~\ref{fig:referentialOfTheBeam} of this document, we plot diagrams that illustrate the evolution of the static equilibria of the strip as a function of boundary actuation. We quantify changes in the strip's equilibrium states by tracking the transverse deflection of a single vertex of  the infinite-dimensional strip: to highlight the evolution of the initially-stable pair of equilibria U\textsubscript{A,B} and the initially-unstable pair S\textsubscript{A,B}, we plot the midpoint deflection as a function of the bifurcation parameter $\mu_d$ for translational actuation and as a function of $\mu$ for rotational actuation.   
On each diagram, we show two sets of results: results obtained based on numerical simulations of the nonlinear Cosserat rod theory and results based on the quasilinear Euler beam model.

%The diagrams in Figs.~1 and~3 of the main text and Figs.~\ref{fig:forwardBackwardSW} and~\ref{fig:forwardBackwardGMV} of this document are pseudo-bifurcation diagrams. 
To identify without doubt the nature of the bifurcation in each system, we conduct rigorous asymptotic analysis (see companion paper~\cite{radisson2022PRE}) that leads to reduced normal forms (eqs. (4) and (5) in the main text) describing the nature of the underlying bifurcation. These equations reduce the strip dynamics to a one degree of freedom system in the vicinity of the bifurcation. In Fig 2 of the main text, the bifurcation diagrams associated with the reduced forms Eqns. (4) and (5) of the main text are compared to: (i) data obtained from the analysis of the Euler beam model, (ii) numerical data obtained from solving the discrete Cosserat equations, and (iii) experimental data obtained in \cite{gomez2017}. In the following, we describe the method we employed to plot these diagrams.

\bigskip
\par\noindent
\textbf{A. Reduced forms.}
The equilibrium points $A_\textrm{eq}$ are obtained by solving the stationnary version of Eqns. (4) and (5) of the main text. Results are plotted as a function of $\Delta \mu$ in Fig. 2G-I of the main text. Their stability is assessed through standard stability analysis of the equilibria $A_\textrm{eq}$ according to the dynamics described by the reduced equations. Stable branches are depicted in black solid lines and unstable branches in black dashed lines. See \cite{radisson2022PRE} for details.

\bigskip
\par\noindent
\textbf{B. Euler beam model.} These bifurcation diagrams are compared to the equilibrium solutions obtained from the static analysis of the geometrically constrained Euler beam model. In the case of asymmetric boundary actuation, the equilibrium amplitudes $A_{\textrm{eq}_1}$ and $A_{\textrm{eq}_2}$ of (4) are to be compared to 
amplitudes associated with the S\textsubscript{B} and U\textsubscript{A} shapes of the elastic strip. 
In the case of symmetric boundary actuation, the equilibrium amplitudes $A_{\textrm{eq}_1}$, $A_{\textrm{eq}_2}$, and $A_{\textrm{eq}_3}$ of (5) are to be compared to amplitudes associated with the U\textsubscript{A}, S\textsubscript{B} and S\textsubscript{A} shapes of the elastic strip. Lastly, for the antisymmetric boundary actuation, the amplitudes $A_{\textrm{eq}_1}$, $A_{\textrm{eq}_2}$, and $A_{\textrm{eq}_3}$ of (5) are to compared to the 
amplitudes associated with S\textsubscript{B}, U\textsubscript{B} and U\textsubscript{A} shapes
of the elastic strip. To calculate the amplitude associated with the equilibria of the elastic strip, we use the approximation
%------
\begin{equation}
	A_{\textrm{eq}}\approx\frac{W_{\textrm{eq}}(X)-W_{\textrm{eq}}^*(X)}{\Phi_0(X)}
	\label{eq:A_approx}
\end{equation}
%----------
For each equilibrium (except U\textsubscript{A} in the symmetric case and S\textsubscript{B} in the antisymmetric case (see below)), this amplitude is plotted for the mid-point of the strip ($X=0$) in the case of the asymmetric and antisymmetric cases and for the quarter-point ($X=1/4$) in the symmetric case. They converge in each case to the equilibria of (4) and (5) (black lines) in the limit $\Delta\mu\ll 1$. We note that as $\Phi_0$ is the mode that the strip follows at leading order to go away from its bifurcation shape, this amplitude can be plotted for any value of $X$ and will always converge to the equilibria of the amplitude equations in the very vicinity of the bifurcation.

For the U\textsubscript{A} shape in the symmetric case and the S\textsubscript{B} shape in the antisymmetric case, it is shown by a scaling analysis (see \cite{radisson2022PRE}) that the largest component in the way these modes go away from the shape at bifurcation is of order $O(\Delta\mu)$. Therefore, for $\Delta\mu\ll1$, there is nothing as large as $O(\Delta\mu^{1/2})$ (leading order) in these data and $A$ is simply set to zero.

All the obtained branches are plotted in Fig. 2  in green for U\textsubscript{A} and U\textsubscript{B} and brown for S\textsubscript{A} and S\textsubscript{B} with dashed (respectively full) lines for unstable (respectively stable) equilibria.

\bigskip
\par\noindent
\textbf{C. Numerical simulations based on the 3D Cosserat rod theory.} The reduced amplitude $A_\textrm{eq}$ of the equilibrium shapes that are computed numerically is obtained following the exact same procedure as the one described in the previous section although obviously only the stable solutions are obtained in these numerical simulation. The obtained branches of static solutions are plotted in Fig. 2  using green square symbols.

\bigskip
\par\noindent
\textbf{D. Experimental data of Gomez et al.}
We compare the bifurcation diagram in the case of asymmetric boundary actuation in Fig. 2G to experimental data obtained by Gomez et al. \cite{gomez2017}. The amplitude $A_{\textrm{eq}}$ from their data is obtained from the truncated expansion of the equilibrium shapes $W_\textrm{eq}(X)$ as done for the data based on the Euler beam model and Cosserat rod theory. 

\bigskip
\par\noindent
\textbf{E. Comment on the location of the bifurcation point.} 
In plotting data from the Euler beam model, from numerical simulations of the 3D Cosserat rod theory, and from the experiments of Gomez et al. \cite{gomez2017}, we shifted the data in each case to place the corresponding bifurcation point at $\Delta\mu=0$. Indeed, each set of data predicts a slightly different value of the bifurcation threshold $\mu^\ast$. In~\cite{gomez2017}, the actual bifurcation point associated with each set of measurements was taken from a parabolic fit of their measurement in the vicinity of the transition (see \cite{gomez2017} Supplemental Material). 

Shifting the data in Fig. 2 of the main text so that the corresponding bifurcation point is at $\Delta\mu=0$ allows us to focus on the way the strip goes away from its configuration at the bifurcation point and not on the quality of the prediction of the bifurcation point itself. It is obvious that the Euler beam framework is valid only in the small deflection limit and that the prediction of the bifurcation point will therefore not be valid for large $\Delta L$ values. Our numerical simulations show that, although the actual bifurcation point goes away from that predicted by the Euler beam model when $\Delta L$ is increased, the behavior predicted in the vicinity of the bifurcation is robust as long as the distance $\Delta\mu$ is taken from the corresponding bifurcation point. This is also true for the experimental data of \cite{gomez2017}. Although they observe variations of the position of the bifurcation point (see Fig. S2 in their supplemental document), the way the strip goes away from its bifurcation shape seems to be in good agreement with the predictions based on the Euler beam model and Cosserat rod theory. Indeed, although the prediction of the exact value of the bifurcation point may not be reliable as $\Delta L$ is increased, we expect the nature of the bifurcation predicted by our analysis, and therefore the time and spatial scaling around the bifurcation, to be robust as they are dictated by rules of symmetry only (see main text). For a reliable prediction of the bifurcation point a fully non-linear analysis as the one carried out by Sano \& Wada \cite{sano2018} should be used instead (Fig. 3B in their paper shows how the prediction obtained in the small deflection limit goes away from the actual value when the longitudinal strain is increased).

\section{Methods for plotting energy landscapes}
\label{sec:energyplot}

The  energy landscape plotted in Fig. 4 of the main text  is a simplified version of the  actual energy landscape of the system at $\mu=0$. It exhibits two potential wells corresponding to the two fundamental Euler buckling modes U\textsubscript{A} and U\textsubscript{B} and two lowest energy barriers corresponding to the unstable equilibria S\textsubscript{A} and S\textsubscript{B} that prevent the system from switching freely from U\textsubscript{A} to U\textsubscript{B} and vice versa.

To obtain this simplified energy landscape, the bending energy $\mathcal{E}_b=EI/2\int_{-1/2}^{1/2}w_{xx}^2dx$ of each static equilibrium obtained from the static analysis of the beam equation are plotted in a 2D space spanned by $W_S(0)$ the midpoint deflection of the symmetric modes (U, W) and $W_A(-1/4)$ the deflection of the antisymmetric modes of buckling (S) at $X=-1/4$. By convention, the antisymmetric component of the symmetric modes is set to zero and vice versa. The two fundamental modes U\textsubscript{A} and U\textsubscript{B} have opposite values in $W(0)$ this is due to the fact that they are the symmetric of each other by the transformation $X\rightarrow -X$ $W\rightarrow-W$. On the energy landscape they are therefore symmetrically distributed around the direction $W_A(1/4)$ and have the same bending energy. In the same way, S\textsubscript{A} and S\textsubscript{B} have opposite values in $W(1/4)$ this is because they are the symmetric of each other by the transformation $X\rightarrow -X$. They are also energetically equivalent but their bending energy is higher than U\textsubscript{A} and U\textsubscript{B}. Finally W\textsubscript{A} and W\textsubscript{B} are symmetric and have therefore a zero antisymmetric component ($W_A(1/4)=0$). In addition, their midpoint deflection is actually zero so their component $W_S(0)$ is also zero and they both lie at the origin of this 2D space. Their bending energy is even higher than the two S modes. Even, in the simple space adopted here, the energy landscape is not as simple because the higher harmonics of buckling constitutes other energy bumps on both axis depending on whether they are symmetric or antisymmetric. Here we have represented only the three first harmonics. However, all the harmonics that are not represented correspond to higher bending energy states and the two $S$ shapes constitutes therefore the two lowest energy barriers preventing the system to transit from U\textsubscript{A} to U\textsubscript{B} or vice versa. 

\begin{figure}[!t]
	\centering
	\includegraphics[width =\textwidth]{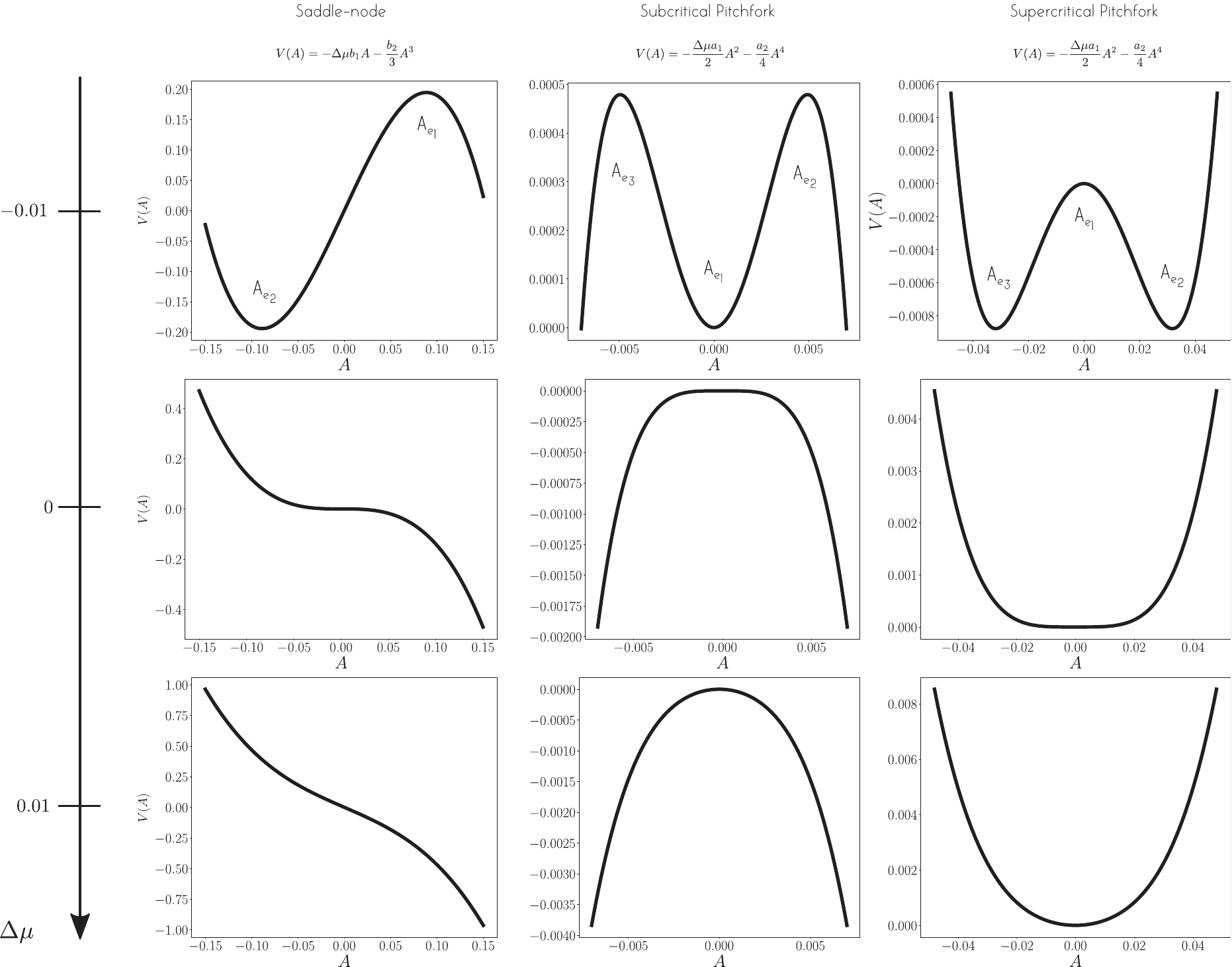}
	\caption{{\textbf{Energy landscapes.}  Plot of the potential landscape $V(A)$ associated with \eqref{eq:saddleNodeFinalForm}, \eqref{eq:subcriticalPitchforkFinalForm} and \eqref{eq:supercriticalPitchforkFinalForm} for each actuation. In each case, the potential landscape is represented for three values of $\Delta\mu$ with one right before the bifurcation ($\Delta\mu<0$), one at the bifurcation ($\Delta\mu=0$), and one after the bifurcation ($\Delta\mu>0$)}.  Equilibria are indicated. The evolution of the potential landscape corresponds in each case to the one plotted schematically on Fig.~4 of the main paper.}
	\label{fig:potentialLandscapesAsymptotic}
\end{figure}

When the boundaries of the strip are rotated, this standard energy landscape is reshaped until one (or both) of the two lowest energy barriers ``breaks," thus allowing the system to transition from one state to another. The deformed energy landscapes are plotted in Fig. 4 of the main paper for the three types of rotational actuation: asymmetric, symmetric, and antisymmetric. The representation of these energy landscapes in Fig. 4 is semi-schematic: we compute the energy values at the equilibria in a rigorous manner but the shape of the energy landscape between two consecutive equilibria is schematic. This semi-schematic representation is not completely ad-hoc. We know from our asymptotic analysis and the reduced normal forms that the energy landscape in the vicinity of the bifurcation has the actual shape of the potential landscape associated with a saddle-node, subcritical pitchfork and supercritical pitchfork for the asymmetric, symmetric and antisymmetric actuation, respectively. The potential landscapes associated with the reduced forms are plotted in Fig. \ref{fig:potentialLandscapesAsymptotic} for three distinct values of the bifurcation parameter $\Delta \mu = \mu - \mu^\ast$. They form the backbone based on which we built the complete, although simplified, energy landscapes in Fig. 4 of the main text.

\section{Tapered strip}

Our work provides tools to predict the type of shape transition an elastic structure is likely to undergo by examining the geometric symmetries of the system. This allows us to design systems that achieve a desired kind of shape transition. The goal of this section is to provide the reader with an example of how these findings can be applied.

Let's say we want to design a system where a buckled elastic strip clamped at both ends undergoes a non-linear snap-through, via a saddle-node bifurcation, while being actuated by antisymmetric rotation of its boundaries. To obtain a saddle-node bifurcation, the actuation needs to break the twin symmetries of both the fundamental U-modes and the S-modes which constitute the  lowest energy barriers. 
Our study showed that, for a geometrically homogeneous strip, antisymmetric actuation results in a supercritical pitchfork bifurcation because, although it breaks the twin symmetry of the S-modes, it satisfies the twin symmetry of the U-modes: antisymmetric boundary actuation breaks the S-twin symmetry (left-right transformation $x\rightarrow-x$) and maintains the U-twin symmetry ($\pi$-rotation $x\rightarrow-x$ and $w\rightarrow-w$). To achieve a saddle-node bifurcation with this actuation, we must break the U-twin symmetry. This can be achieved by simply using a tapered strip instead of a homogeneous strip. By tapering the strip cross-sectional area, the $x \rightarrow -x$ symmetry is suppressed and with it both the U-twin and S-twin symmetries are also suppressed. 

We assess these predictions using numerical simulations based on the 3D Cosserat rod theory, and analysis based on the geometrically-constrained Euler beam model of an antisymmetrically actuated clamped-clamped strip of thickness $h$ that is linearly decreasing  along its length. All remaining parameters of the strip are the same as the ones adopted throughout this study.
%, except for the thickness $h$ that is linearly decreasing along the curvilinear coordinate $s$ of the strip. 
In Fig. 4 of the main text, we show the effect of the tapering on the two equilibrium shapes U\textsubscript{A} and U\textsubscript{B} when a zero slope ($\mu=0$) is imposed at the boundaries. Specifically,  we decreased the thickness $h$ of the strip linearly from $h=2$ mm at the left end of the strip to $h=1$ mm at the right end of the strip. This strong tapering is chosen to make the left-right symmetry breaking on the equilibrium shapes visually obvious.

We now analyze the evolution of the equilibria U\textsubscript{A} and U\textsubscript{B} when the boundaries are rotated in an antisymmetric fashion. In the following, we study a strip with a tapering going from $h=2$ mm (left) to $h=1.6$ mm (right). This weaker tapering is chosen in order to see how a slight symmetry breaking affects the transition observed for the homogeneous strip. We apply the same methodology employed in this study: we increase quasi-statically the angle $\alpha$ applied at the boundaries until the strip starts to stretch along its length. Contrary to the homogeneous strip, the transition from a bistable system to a system with only one equilibrium is no longer smooth: when a certain value $\mu^\ast \approx 1.6098316$ of the non-dimensional angle applied at the boundaries is reached, the U\textsubscript{A} equilibrium suddenly snaps through to the U\textsubscript{B} equilibrium. This is shown on the bifurcation diagram in Fig. \ref{fig:taperedBeam}B where the data obtained numerically for the tapered strip (green symbols) are compared to the bifurcation diagram obtained analytically for the homogeneous strip (black lines). 

In order to confirm the nature of this shape transition, we analyze how the dynamic slow-down when the strip approaches the transition. We measure the eigenfrequency of the fundamental mode of vibration around the equilibrium U\textsubscript{A} at different distance $\Delta\mu=\mu-\mu^\ast$ from the bifurcation point $\mu^\ast$. For each $\Delta \mu$ value, we start the simulation with the corresponding equilibrium configuration. At $t=0$, we apply a sudden 'kick' to the strip by applying a small point force on one of the vertex of the Cosserat rod. The resulting dynamics is then analyzed by performing a Fourier transform of the dynamic evolution of the midpoint deflection of the strip from which we extract the fundamental frequency (see \cite{radisson2022PRE} for methods). The obtained values are then plotted against $|\Delta\mu|$ in Fig. 4 of the main text (green dots). The analytical solution of the homogeneous strip obtained from the asymptotic analysis in the vicinity of the bifurcation is plotted on the same figure (black line) for comparison. The obtained slowing down follows the scaling $\sqrt{|\sigma^2|}=|\Delta\mu|^{1/4}$ which is to be compared with the scaling obtained for the homogeneous strip $\sqrt{|\sigma^2|}=|\Delta\mu|^{1/2}$ (Fig. 4 of the main text,  black line). These two scalings are known to be the signature of second order in time saddle-node and pitchfork bifurcation, respectively (see main text). Their presence proves that the symmetry breaking due to the tapering of the strip has turned the supercritical pitchfork observed for this actuation in the case of a homogeneous strip into a saddle-node bifurcation.

%%%%%%%%%%%%%%%%%%%%%%%%%%%%

\bibliography{referencesSupplemental}

\newpage

%%%%%%%%%%%%%%%%%%%

\begin{table}
\centering
\caption{Boundary conditions associated with each type of boundary actuation.}
%\bgroup
%\def\arraystretch{2}
%\begin{tabular}{l l} 
\begin{tabular}{p{0.17\textwidth}p{0.83\textwidth}}
\toprule
\multicolumn{2}{c}{$\textbf{EULER BUCKLING}$}
\\
\toprule
$\textbf{Clamped-Hinged:}$ & $\qquad \left.W\right|_{\rm left}=0,  \qquad \left.\dfrac{\partial W}{\partial X}\right|_{\rm left}=0, \qquad
\qquad \ \ \  \left.W\right|_{\rm right}=0, \qquad \left. \dfrac{\partial^2 W}{\partial X^2}\right|_{\rm right}=0\qquad$\\[4ex]
$\textbf{Hinged-Hinged:}$& $\qquad \left.W\right|_{\rm left}=0,  \qquad \left.\dfrac{\partial^2 W}{\partial X^2}\right|_{\rm left}=0, 
\qquad
\qquad \ \ \left.W\right|_{\rm right}=0, \qquad \left. \dfrac{\partial^2 W}{\partial X^2}\right|_{\rm right}=0\qquad$\\[4ex] 
$\textbf{Clamped-Clamped:}$ & $\qquad \left.W\right|_{\rm left}=0,  \qquad \left.\dfrac{\partial W}{\partial X}\right|_{\rm left}=0, \qquad 
\qquad \ \ \ \left.W\right|_{\rm right}=0, \qquad \left. \dfrac{\partial W}{\partial X}\right|_{\rm right}=0\qquad$\\[4ex]
\toprule
\multicolumn{2}{c}{$\textbf{TRANSLATIONAL ACTUATION of EULER BUCKLED STRIP}$}                        \\
\toprule
$\textbf{Clamped-Hinged:} $& $\qquad \left.W\right|_{\rm left}=\mu_d,  \qquad \left.\dfrac{\partial W}{\partial X}\right|_{\rm left}=0, 
\qquad
\qquad \ \left.W\right|_{\rm right}=0,
\qquad \left. \dfrac{\partial^2 W}{\partial X^2}\right|_{\rm right}=0\qquad$\\[4ex]
$\textbf{Hinged-Hinged:} $& $\qquad \left.W\right|_{\rm left}=\mu_d,  \qquad \left.\dfrac{\partial^2 W}{\partial X^2}\right|_{\rm left}=0, 
\qquad \qquad \left.W\right|_{\rm right}=0,
\qquad 
\left. \dfrac{\partial^2 W}{\partial X^2}\right|_{\rm right}=0\qquad$\\[4ex] 
$\textbf{Clamped-Clamped:}$ & $\qquad \left.W\right|_{\rm left}=\mu_d, \qquad \left.\dfrac{\partial W}{\partial X}\right|_{\rm left}=0,
\qquad \qquad \ \left.W\right|_{\rm right}=0, 
\qquad  \
\left. \dfrac{\partial W}{\partial X}\right|_{\rm right}=0\qquad$\\[4ex]
\toprule
\multicolumn{2}{c}{$\textbf{ROTATIONAL ACTUATION of EULER BUCKLED STRIP}$}                        \\
\toprule
$\textbf{Asymmetric:}$ & $\qquad \left.W\right|_{\rm left}=0, \qquad \left.\dfrac{\partial W}{\partial X}\right|_{\rm left}=\mu,
\qquad \qquad \ \ \  \left.W\right|_{\rm right}=0,  \qquad \left. \dfrac{\partial W}{\partial X}\right|_{\rm right}=0\qquad$\\[4ex]
$\textbf{Symmetric:} $& $\qquad \left.W\right|_{\rm left}=0,
\qquad \left.\dfrac{\partial W}{\partial X}\right|_{\rm left}=\mu, 
\qquad \qquad \ \ \  \left.W\right|_{\rm right}=0,  \qquad \left. \dfrac{\partial W}{\partial X}\right|_{\rm right}=-\mu \qquad$\\[4ex] 
$\textbf{Antisymmetric:}$ & $\qquad \left.W\right|_{\rm left}=0,\qquad \left.\dfrac{\partial W}{\partial X}\right|_{\rm left}=\mu,
\qquad \qquad \ \ \  \left.W\right|_{\rm right}=0,  \qquad \left. \dfrac{\partial W}{\partial X}\right|_{\rm right}=\mu\qquad$\\[4ex]
\bottomrule
\end{tabular}
\label{tab:dimensional_boundary_conditions}
\end{table}

%%%%%%%%%%%%%%%%%%%%%%%%%%%%
\begin{table}[!t]
\caption{\textbf{Euler buckling:} Static equilibria of the Euler buckled strip. Here, the origin of the $X$-coordinate is placed at the left end of the strip $(X\in [0,1])$}
\begin{tabular}{p{0.17\textwidth}p{0.83\textwidth}}
\toprule
\multicolumn{2}{c}{\textbf{\textsc{Euler-Buckling: Clamped-Hinged}}}\\
\toprule 
$\textbf{All Solutions}$		       
		      & ${\tan(\Lambda_{n})=\Lambda_{n}}$ \\[2mm]
		       & $W_n(X)=A\sin(\Lambda_n X)+B\cos(\Lambda_n X)+ CX+D$
		       \\[3ex]
		       & {\footnotesize $A= \displaystyle\frac{\pm 2\sqrt{2} }{\sqrt{\Lambda_n \left(2 \Lambda_n^3-\Lambda_n^2 \textrm{S2}-8 \textrm{S1}+\textrm{S2}+8 \Lambda_n \textrm{C1}-2 \Lambda_n \textrm{C2}\right)}}$},
		       \quad
		        {\footnotesize $B=-\Lambda_n A$}, \quad {\footnotesize $C=B$},\quad {\footnotesize $D=-B$}
		       \\[3ex]
		       &{\footnotesize $\textrm{S1}\!=\!\sin\Lambda_{n},~\textrm{C1}\!=\!\cos\Lambda_{n}, \ \textrm{S2}\!=\!\sin2\Lambda_{n},~\textrm{C2}\!=\!\cos2\Lambda_{n}$}
		       \\[3ex]
		       \toprule
 \multicolumn{2}{c}{\textbf{\textsc{Euler-Buckling: Hinged-Hinged}}}\\
\toprule 
$\textbf{All Solutions}$	
			& ${\Lambda_{n} =(n+1)\pi}$  \\[3ex]
			& $\displaystyle W_{n}(X)=\pm \frac{2}{\Lambda_{n}}\sin\left(\Lambda_{n} X\right)$ 
			\\
\toprule
 \multicolumn{2}{c}{\textbf{\textsc{Euler Buckling: Clamped-Clamped}}}\\
\toprule
$\textbf{Even harmonics}$      & ${\Lambda_{2n} = 2(n+1)\pi}$ \\[3ex]
		      	& $W_{2n}(X)=\displaystyle \pm \frac{2}{\Lambda_{2n}}\left(\cos\left(\Lambda_{2n} X\right)-1\right)$
		      \\[4ex]
		      &
		      \\
$\textbf{Odd harmonics}$ & $\displaystyle {\tan(\dfrac{\Lambda_{2n+1}}{2})= \dfrac{\Lambda_{2n+1}}{2}}$ \\[3ex]
		        & $W_{2n+1}(X)=A\sin(\Lambda_{2n+1} X)+B\cos(\Lambda_{2n+1} X)+ CX+D$
		       \\ [3ex]
		       & {\footnotesize $A = \displaystyle \frac{\pm 4\sqrt{2}S1^2}{\sqrt{-\Lambda\left[-2\Lambda_{2n+1}^3 + \Lambda_{2n+1}^2(4\textrm{S1} 
		       		+ \textrm{S2})+4\Lambda_{2n+1}(\textrm{C2}-\textrm{C1})+4\textrm{S1}-2\textrm{S2}\right]}}$}, \\[6ex]
		       		&
		       		{\footnotesize $B=\displaystyle \frac{\Lambda_{2n+1}-\textrm{S1}}{\textrm{C1}-1}A$}, \quad {\footnotesize $C=-\Lambda_{2n+1} A$},\quad  {\footnotesize $D=-B$}
		       \\[3ex]
		       &{\footnotesize $\textrm{S1}\!=\!\sin\Lambda_{2n+1},~\textrm{C1}\!=\!\cos\Lambda_{2n+1}, \ 	\textrm{S2}\!=\!\sin2\Lambda_{2n+1},~\textrm{C2}\!=\!\cos2\Lambda_{2n+1}$}
		       \\[3ex]
			\bottomrule
\end{tabular}
\label{tab:eigenEulerBuckling}
\end{table}

%%%%%%%%%%%%%%%%%%%%%%%%%%%%
\begin{table}[!t]
\caption{\textbf{Translational boundary actuation:} static equilibria of Euler-buckled strip under Translational boundary actuation. Here, the origin of the $X$-coordinate is placed at the left end of the strip $(X\in [0,1])$}
\begin{tabular}{p{0.15\textwidth}p{0.85\textwidth}}
\toprule
\multicolumn{2}{c}{ \textbf{\textsc{Clamped-Hinged}}}\\
\toprule
\textbf{All solutions}		       & 
{ $\displaystyle\frac{4\left(\textrm{S1}-\Lambda_{n}\textrm{C1}\right)^2}{\Lambda_{n}\left(3\Lambda_{n}\textrm{C1}^2) +
		       	\Lambda_n\textrm{S1}^2-3\textrm{S1}\textrm{C1}\right)} = \mu_d^2$} \\[5ex]
		       & $W_n(X)=A\sin(\Lambda_n X)+B\cos(\Lambda_n X)+ CX+D$
		       \\[2ex]
		      &
		   {$\textrm{S1}=\sin\Lambda_{n},\quad\textrm{C1}=\cos\Lambda_{n}$}, \quad {$A= \displaystyle \frac{\mu_d \textrm{C1}}{\Lambda_n \textrm{C1}-\textrm{S1}}, \quad B=\displaystyle \frac{-\mu_d \textrm{S1}}{\Lambda_n \textrm{C1}-\textrm{S1}}$}
		      \quad {$C=
		      - \Lambda_n A, \quad D=\Lambda_n A$}
		       \\[3ex]
		        \toprule
\multicolumn{2}{c}{ \textbf{\textsc{Hinged-Hinged}}}\\
\toprule
\textbf{All solutions}			& ${\Lambda_{n} = (n+1)\pi}$ \\[2ex]
			 & $W_{n}(X)=\displaystyle\pm \frac{\sqrt{4-2\mu_d^2}}{\Lambda_{n}}\sin\left(\Lambda_{n} X\right)+\mu_d(1-X)$
			\\[3ex]

	\toprule
\multicolumn{2}{c}{ \textbf{\textsc{Clamped-Clamped}}}\\
\toprule
$\textbf{Even harmonics}$      
& 
${\Lambda_{2n} = 2(n+1)\pi}$  \\[2ex]
& $W_{2n}(X) = \displaystyle \pm\frac{\sqrt{4-3\lambda^2}}{\Lambda_{2n}}\left(\cos(\Lambda_{2n} X)-1\right)+\lambda\left(\frac{\sin(\Lambda_{2n}X)}{\Lambda_{2n}}-X+1\right)$ \\
& \\
$\textbf{Odd harmonics}$ & 	
$\displaystyle \frac{\left(2\Lambda_{2n+1}\textrm{C2}-4\textrm{S2}\right)^2}{\Lambda_{2n+1}\left(2\Lambda_{2n+1} +\Lambda_{2n+1}\textrm{C1}-3\textrm{S1}\right)} = \mu_d^2$ \\[3ex]
& 	$W_{2n+1}(X)=A\sin(\Lambda_{2n+1} X)+B\cos(\Lambda_{2n+1} X)+ CX+D$
		       \\[3ex]
		       &
		       { $\textrm{S1}\!=\!\sin\Lambda_{2n+1},\quad \textrm{C1}\!=\!\cos\Lambda_{2n+1}$}, \quad 
		       { $\textrm{S2}\!=\!\sin\dfrac{\Lambda_{2n+1}}{2},\quad
		       \textrm{C2}\!=\!\cos\dfrac{\Lambda_{2n+1}}{2}$}
		   \\[3ex]
		       & 	 $A=\displaystyle \frac{-\mu_d \textrm{S1}}{2-2\textrm{C1}-\Lambda_{2n+1}\textrm{S1}}$,
		       \qquad  $B=\displaystyle\frac{-\mu_d (\textrm{C1}-1)}{2-2\textrm{C1}-\Lambda_{2n+1}\textrm{S1}}$
		       \\[3ex]
		   & { $C=\displaystyle\frac{\mu_d\Lambda_{2n+1} \textrm{S1}}{2-2\textrm{C1}-\Lambda_{2n+1}\textrm{S1}}$}, \qquad
		   {$D=\displaystyle\frac{\mu_d(1-\textrm{C1}-\Lambda_{2n+1}\textrm{S1})}{2-2\textrm{C1}-\Lambda_{2n+1}\textrm{S1}}$}
		       \\
		       \\
			\bottomrule
\end{tabular}
\label{tab:eigenTranslational}
\end{table}

%%%%%%%%%%%%%%%%%%%%%%%%%%%%
\begin{table}[!t]
\caption{\textbf{Rotational boundary actuation:} static equilibria of clamped-clamped buckled strip under rotational actuation. Here, the origin of the $X$-coordinate is placed at the left end of the strip $(X\in [0,1])$}
\begin{tabular}{p{0.15\textwidth}p{0.85\textwidth}}
\toprule
\multicolumn{2}{c}{\textbf{\textsc{Asymmetric Actuation}}}\\
\toprule
%\multicolumn{3}{l}{\cellcolor[rgb]{0.85,0.85,0.85} \textbf{Asymmetric}} \\ %\midrule
$\textbf{All Solutions}$    &  $\displaystyle \frac{8\Lambda_n(\Lambda_n\textrm{S1}+2\textrm{C1}-2)^2}{2\Lambda_n^3-\Lambda_n^2\left(4\textrm{S1}
							+\textrm{S2}\right)+4\Lambda_n\left(\textrm{C1}-\textrm{C2}+2\textrm{S2}-4\textrm{S1}\right)}=\mu^2$
		      \\[5ex]
		      &	 $W_{n}(X) = \displaystyle A\sin(\Lambda_n X)+B\cos(\Lambda_n X)+ CX+D$ 
		      \\[3ex]
		      & {\footnotesize $A=\displaystyle \frac{\mu(\textrm{C1}+\Lambda_n\textrm{S1}-1)}{\Lambda_n(\Lambda_n\textrm{S1}+2\textrm{C1}-2)},\quad B=\displaystyle \frac{\mu(\Lambda_n\textrm{C1}-\textrm{S1})}{\Lambda_n(\Lambda_n\textrm{S1}+2\textrm{C1}-2)},$ \quad $C=\displaystyle \frac{-\mu\textrm{S3}}{\Lambda_n\textrm{C3}-2\textrm{S3}},\quad D=\displaystyle \frac{-\mu(\Lambda_n\textrm{C1}-\textrm{S1}).}{\Lambda_n(\Lambda_n\textrm{S1}+2\textrm{C1}-2)}$}
		      \\[3ex]
		       &{\footnotesize $\textrm{S1}\!=\!\sin\Lambda_{n},\quad\textrm{C1}\!=\!\cos\Lambda_{n},\quad\textrm{S2}\!=\!\sin(2\Lambda_{n}),\quad\textrm{C2}\!=\!\cos(2\Lambda_{n}),\quad \textrm{S3}\!=\!\sin(\Lambda_{n}/2),~\textrm{C3}\!=\!\cos(\Lambda_{n}/2).$}
		       \\[3ex]
                      \toprule
\multicolumn{2}{c}{\textbf{\textsc{Symmetric Actuation}}}\\
\toprule
%\multicolumn{3}{l}{\cellcolor[rgb]{0.85,0.85,0.85} \textbf{Symmetric}} \\ %\midrule 
$\textbf{Even harmonics}$	& $\displaystyle\dfrac{2\Lambda_{2n}(1-\cos(\Lambda_{2n}))}{\Lambda_{2n} - \sin(\Lambda_{2n})} = \mu^2$\\[5ex]		       
			& $ \displaystyle W_{2n}(X)=A\sin(\Lambda_{2n} X)+B\cos(\Lambda_{2n} X)+ CX+D$ 
			\\ [3ex]
			& {\footnotesize $A=\displaystyle\frac{\mu}{\Lambda_{2n}},\quad B=\displaystyle\frac{\mu\cot(\Lambda_{2n}/2)}{\Lambda_{2n}},\quad C=0,\quad D=-B$}
			\\[3ex]
$\textbf{Odd harmonics}$	& $\tan(\dfrac{\Lambda_{2n+1}}{2})\!=\!\dfrac{\Lambda_{2n+1}}{2}$	\\[3ex]
			& $W_{2n+1}(X)=A\sin(\Lambda_{2n+1} X)+B\cos(\Lambda_{2n+1} X)+ CX+D$
			\\[3ex]
			& 	{\footnotesize $A= K_1\pm K_2$, \quad $B=\displaystyle A\frac{\textrm{C1}-1}{\textrm{S1}} + \frac{2\mu}{ \Lambda_{2n+1}\textrm{S1}},\quad C=\mu-\Lambda_{2n+1} A,\quad D=-B$}
			\\[3ex]
			&	{\footnotesize $K_1 = \displaystyle \frac{\mu}{\Lambda_{2n+1}}$, \quad $K_2 = \displaystyle \frac{2\sqrt{4-\mu^2}}{\Lambda_{2n+1}^2},\quad\textrm{S1}\!=\!\sin\Lambda_{2n+1},\quad\textrm{C1}\!=\!\cos\Lambda_{2n+1}.$}\\[3ex]
\toprule
\multicolumn{2}{c}{\textbf{\textsc{Antisymmetric Actuation}}}\\
\toprule
%\multicolumn{3}{l}{\cellcolor[rgb]{0.85,0.85,0.85}\textbf{Antisymmetric}}
			\\ %\midrule
$\textbf{Even harmonics}$	&  	${\Lambda_{2n} = 2(n+1)\pi}$ \\[3ex]
		 	& 	 $\displaystyle W_{2n}(X)=\pm\frac{\sqrt{4-\mu^2}}{\Lambda_{2n}}\left(\cos(\Lambda_{2n} X)-1\right)
					+\mu\left(\frac{\sin(\Lambda_{2n}X)}{\Lambda_{2n}}\right)$	
			\\[3ex]
$\textbf{Odd harmonics}$   & 	{ $\displaystyle \frac{4\left(\Lambda_{2n+1}\textrm{C2}-2\textrm{S2}\right)^2}{\Lambda_{2n+1}^2 
					+ \Lambda_{2n+1} \textrm{S1}+4\textrm{C1}-4} =\mu^2$}\\[5ex]
			&	$W_{n}(X)=A\sin(\Lambda_{2n+1} X)+B\cos(\Lambda_{2n+1} X)+ CX+D$
			\\[3ex]
			&  {\footnotesize $A= \displaystyle\frac{\mu \textrm{S1}}{\Lambda_{2n+1} \textrm{S1}+2 \textrm{C1}-2}, \quad B=\displaystyle-\frac{\mu \textrm{S2}}{\Lambda_{2n+1}\textrm{C2}-2 S2},\quad C =\displaystyle\frac{2 \mu (\textrm{C1}-1)}{\Lambda_{2n+1} \textrm{S1}+2 \textrm{C1}-2}, \quad D=\displaystyle\frac{\mu \textrm{S2}}{\Lambda_{2n+1} \textrm{C2}-2 \textrm{S2}}.$}
			\\[4ex]
		       &{\footnotesize $\textrm{S1}\!=\!\sin\Lambda_{2n+1},\quad\textrm{C1}\!=\!\cos\Lambda_{2n+1}, \quad\ 
						\textrm{S2}\!=\!\sin(\Lambda_{2n+1}/2),\quad\textrm{C2}\!=\!\cos(\Lambda_{2n+1}/2).$}
		\\[3ex]	
		    \bottomrule
\end{tabular}
\label{tab:eigenRotational}
\end{table}